\RequirePackage{fix-cm}
\documentclass[smallextended]{svjour3}       
\smartqed  
\usepackage{graphicx}
\usepackage{enumerate}
\usepackage{amsmath}
\usepackage{hyperref}
\usepackage{feynmp}
\usepackage{slashed}
\usepackage{array}
\newcommand{\bq}{\begin{equation}}
\newcommand{\eq}{\end{equation}}
\newcommand{\bqn}{\begin{eqnarray}}
\newcommand{\eqn}{\end{eqnarray}}
\newcommand{\nb}{\nonumber}
\newcommand{\lb}{\label}
\newcommand{\e}{\langle e \rangle}
\newcommand{\A}{\langle A \rangle}
\newcommand{\marrow}[5]{%
    \fmfcmd{style_def marrow#1
    expr p = drawarrow subpath (1/4, 3/4) of p shifted 6 #2 withpen pencircle scaled 0.4;
    label.#3(btex #4 etex, point 0.5 of p shifted 6 #2);
    enddef;}
    \fmf{marrow#1,tension=0}{#5}
}

\begin{document}

\title{A Non-Geometrodynamic Quantum Yang-Mills Theory of Gravity Based on the Homogeneous Lorentz Group
}

\titlerunning{Non-Geometrodynamic Lorentz Yang-Mills Theory of Gravity}        

\author{Ahmad Borzou   
}


\institute{A. Borzou \at
              EUCOS-CASPER, Physics Department, Baylor University, Waco, TX 76798-7316, USA \\
              \email{ahmad\_borzou@baylor.edu}           
}

\date{Received: date / Accepted: date}

\maketitle

\begin{abstract}
In this paper, we present a non-geometrodynamic quantum Yang-Mills theory of gravity based on the homogeneous Lorentz group within the general framework of the Poincare gauge theories. 
The obstacles of this treatment are that first, on the one hand, the gauge group that is available for this purpose is non-compact. On the other hand, Yang-Mills theories with non-compact groups are rarely healthy, and only a few instances exist in the literature.  Second, it is not clear how the direct observations of space-time waves can be explained when space-time has no dynamics. 
We show that the theory is unitary and is renormalizable to the one-loop perturbation.
Although in our proposal, gravity is not associated with any elementary particle analogous to the graviton, classical helicity-two space-time waves are explained. Five essential exact solutions to the field equations of our proposal are presented as well. We also discuss a few experimental tests that can falsify the presented Yang-Mills theory. 
\end{abstract}

\section{Introduction}
\label{sec:intro}
The attempts to describe gravity as a gauge theory started with the seminal work of Utiyama \cite{1956PhRv..101.1597U}, who used the gauge fields of the homogeneous Lorentz group to derive the equations of general relativity (GR). Later Sciama \cite{1964RvMP...36..463S} and Kibble \cite{1961JMP.....2..212K} extended this approach by introducing two possible independent fields that describe the dynamics of a more general geometry than the pseudo-Riemannian of GR, i.e. Poincare gauge theories (PGT). 

Rosen and later Gupta tried to give a non-geometric interpretation to Einstein's equations. Although in the light of the modern tests of gravity such interpretation is unlikely, 
this program has been developed into a method that can derive Yang-Mills theories starting from a linear field equation and by adding the interactions using consistency conditions \cite{1979JMP....20.2264F}.


In this paper, we show that the framework of PGT allows for non-geometrodynamic Yang-Mills description of gravity such that the space-time observable quantities are explained by gravitational fields that live on space-time but distinct from it. 
PGT consist of space-time and tangent spaces. In general, when torsion is not zero, both of the spaces are dynamical. If torsion is set to zero right from the beginning, we still have two conserved currents. Nevertheless, only one of the two spaces can have dynamics. It is often assumed that space-time is the dynamical space in the torsion-free cases, and the tangent spaces are regulated using the tetrad postulate together with the dynamics of space-time. In this paper, we alternatively propose that in the torsion-free scenarios, the tangent space should be taken dynamical, and space-time should be determined using the tetrad-postulate together with the dynamics of the tangent-spaces. 
Therefore, in our scenario, the observable effects of space-time are induced by a dynamical gravitational field that lives on space-time but distinct from it. 
In quantum physics, an indication that an observable has a quantum nature is that it satisfies a Poisson commutation relation in the classical regime. Since in our approach, unlike the geometrodynamic treatments, the space-time metric is not driven by such Poisson bracket, we can assume that it has no quantum nature without violating the principle of quantum mechanics. As a result, space-time remains a classical background even at the smallest distances. 

However, we need to address a few challenges. 
To give dynamics to the tangent spaces, we need to build a Yang-Mills theory based on the homogeneous Lorentz group which is not compact. 
The internal metric corresponding with this group is not positive-definite, and the energies of three of the corresponding gauge fields are not bound from below. If these gauge fields represent physical particles, the theory is unacceptable. 
Fortunately, the homogeneous Lorentz group of the tangent spaces is the Lorentz group of physical observers. As a result, its gauge fields are not tensors of the Lorentz group of physical observers and do not represent elementary particles. Therefore, in our proposal, there exists no massless elementary particle that plays the role of the graviton in GR.
While this prediction does not violate any of the observations and is allowed, we still need to explain the observed space-time waves that travel with the speed of light \cite{Monitor:2017mdv}. So far, to our knowledge, this observation is solely explained through the existence of elementary particles.
In this paper, we present an alternative explanation. We show that even if space-time does not have dynamics of its own, the tetrad postulate can still explain the observed space-time waves by coupling the dynamics of the non-geometrical and non-physical Lorentz gauge fields to the space-time.

In this paper, 
we introduce Quantum Lorentz gauge theory of gravity (QLGT), derive its Feynman rules in the path integral formalism, and calculate the irreducible one-loop diagrams of the theory and show that all of the infinities can be absorbed into its available parameters. Therefore, QLGT is renormalizable to the first loop. We also demonstrate that QLGT is unitary.   
We discuss the classical field equations of Lorentz gauge theory of gravity (LGT) and show that a contraction of the field equations is the same as the divergence of Einstein equations. 
Using a plane wave analysis, we demonstrate that the dynamics of LGT in the tangent spaces induce space-time waves. We show that even though in GR space-time has its dynamics and in LGT it does not, the helicity-2 physical modes of the space-time waves are the same in both GR and LGT.
Finally, we show that the Kerr space-time, the Schwarzschild space-time in the vacuum, the Schwarzschild space-time inside the stars, the de Sitter space-time, the early universe space-time, and the space-time of the matter-dominated universe are the solutions of LGT field equations to all orders of perturbations.

The structure of this paper is as follows. 
An overview of the general framework of the Poincare gauge theories is presented in section~\ref{sec:Symmetry}. 
We construct and quantize our non-geometrodynamic Yang-Mills theory of gravity in section~\ref{sec:QLGT}. In the same section, we discuss that the theory is unitary and is renormalizable to one-loop perturbation. 
The classical field equations, a plane wave analysis, and a few exact solutions of LGT
are presented in section~\ref{sec:Classic}. In section~\ref{Sec:Experiment}, we discuss a few possible experimental tests of the theory. A conclusion is drawn in section~\ref{sec:conclusion}.

\section{An overview of the Poincare gauge theories of gravity in pseudo Riemannian geometry }
\lb{sec:Symmetry}

This section is a summary of the literature on PGT with an emphasis on the aspects that are utilized later in the paper. For in-depth reviews, we refer the reader to \cite{2018IJGMM..1540005O,2006gr.qc.....6062T,2011CQGra..28u5017B,2015CQGra..32e5012K,1996CQGra..13..681W}.
Since the four-dimensional general linear group has no representation that transforms like a spinor under the Lorentz group, we have to define the spinor fields in the tangent spaces on space-time \cite{1972gcpa.book.....W}. 
A set of four orthogonal vectors $e_i$ at every point of space-time defines the tangent spaces, i.e., the Lorentz frames. Here and in the rest of the paper, the Latin indices run from 0 to 3 and refer to the Lorentz frames. The orthonormality of the tetrad indicates that the metric of the tangent spaces is always Minkowskian $\eta^{ij}\equiv e_i \cdot e_j$. The Latin indices are lowered and raised by this Minkowski metric. The components of the tetrad in the space-time coordinate system $e_{i\mu}\equiv e_i\cdot e_{\mu}$ define the space-time metric
\bqn
\lb{Eq:MetricInTermOfTetrad}
g_{\mu\nu} = \eta^{ij}e_{i\mu}e_{j\nu},
\eqn 
where the Greek indices, referring to the coordinates, run from 0 to 3 and are raised and lowered by this latter metric. 

Fermionic fields are the scalars of space-time but the spinors of the tangent spaces
\bqn
\lb{Eq:SpinorTransformation}
&&\tilde{\psi}(x) = \exp\left(\frac{g}{2}S^{ij}\omega_{ij}\right)\psi(x),\nb\\
&&\psi'(x') = \psi(x). 
\eqn
Here we present a Lorentz transformation of the tetrad by a tilde and a coordinate transformation by a prime. Moreover, $g$ is a dimensionless coupling constant, $\omega_{ij}$ is an arbitrary anti-symmetric tensor, and $S^{ij}$ are the six generators of the homogeneous Lorentz group in terms of the commutator of the Dirac matrices
\bqn
S^{ij} = \frac{1}{4}\left[\gamma^i,\gamma^j\right].
\eqn

The tetrad are the vectors of the tangent spaces but the scalars of space-time
\bqn
&&\tilde{e}_i(x) = \Lambda^j_{~i}e_j(x),\nb\\
&&e'_i(x') = e_i(x),
\eqn
where $\Lambda^j_{~i}$ represent the homogeneous Lorentz transformations in the vector space. If the parameter $\omega_{ij}$ in Eq.~\ref{Eq:SpinorTransformation} is very smaller than unity, this reads $\Lambda^j_{~i} \simeq \delta^j_i+\omega^j_{~i}$.

The photon, gluon, W, and Z particles are all vectors of coordinate system, but the scalars of the tangent spaces
\bqn
&&\tilde{A}_{\mu}(x) = A_{\mu}(x),\nb\\
&&A^{'}_{\mu}(x') = \frac{\partial x^{\nu}}{\partial x^{'\mu}}A_{\nu}(x),
\eqn
where the internal index of the fields is not shown. 

If space-time is flat and the two types of transformations are not position-dependent, the Dirac Lagrangian reads
\bqn
{\cal{L}}_{\text{D}} = 
\frac{i}{2}e_i^{~\mu}\bar{\psi}\gamma^i\partial_{\mu}\psi
-
\frac{i}{2}e_i^{~\mu}\bar{\psi}\overleftarrow{\partial}_{\mu}\gamma^i \psi
-
m\bar{\psi}\psi.
\eqn

If the transformations are position-dependent, the partial derivatives should be replaced by the covariant derivatives, such that the derivatives of the fields transform as before. 
For spinor fields, the covariant derivatives read
\bqn
&&D_{\mu}\psi = \left(\partial_{\mu}-\frac{1}{2}gA_{ij\mu}S^{ij}\right)\psi,\nb\\
&&\bar{\psi}\overleftarrow{D}_{\mu}=\bar{\psi}\left(\overleftarrow{\partial}_{\mu}+\frac{1}{2}gA_{ij\mu}S^{ij}\right),
\eqn
where $A_{ij\mu}$ is the gauge field of the homogeneous Lorentz group and is anti-symmetric in the two Latin indices. It transforms as 
\bqn
\lb{Eq:GaugeFieldGeneralTransform}
&&\tilde{A}_{ij\mu}(x) = \Lambda_i^{~m}\Lambda_j^{~n}A_{mn\mu}(x) + 
\partial_{\mu}\Lambda_i^{~l}\Lambda_{jl},\nb\\
&&A^{'}_{ij\mu}(x') = \frac{\partial x^{\nu}}{\partial x^{'\mu}}A_{ij\nu}(x).
\eqn
The first equation indicates that the Lorentz gauge field is not a tensor of the Lorentz group of observers
and therefore possess no observer-independent property like the spin and cannot represent an elementary particle. 

To find the covarinat derivative of the Lorentz vectors, we use the product rule to take the derivative of $B^i \equiv \bar{\psi}\gamma^i\psi$ and use the following identity
\bqn
\left[\gamma^i,S^{mn}\right] = \eta^{im}\gamma^n-\eta^{in}\gamma^m.
\eqn
With a straightforward calculation, one can show that the covariant derivative of a vector of the tangent spaces reads 
\bqn
D_{\mu}B^i = \partial_{\mu}B^i-gA^i_{~j\mu}B^j.
\eqn

So far, we have discussed the covariant derivatives of space-time scalars. If a field transforms like a tensor under coordinate transformations, its covariant derivative should contain space-time connections and reads
\bqn
D_{\mu}B^{i_1i_2\cdots\alpha}_{j_1j_2\cdots\beta} &=& \partial_{\mu}B^{i_1i_2\cdots\alpha}_{j_1j_2\cdots\beta}\nb\\ 
&-& 
gA^{i_1}_{~k\mu}B^{ki_2\cdots\alpha}_{j_1j_2\cdots\beta}-
gA^{i_2}_{~k\mu}B^{i_1k\cdots\alpha}_{j_1j_2\cdots\beta}-\cdots\nb\\
&-& 
gA^{~~k}_{j_1~\mu}B^{i_1i_2\cdots\alpha}_{kj_2\cdots\beta}-
gA^{~~k}_{j_2~\mu}B^{i_1i_2\cdots\alpha}_{j_1k\cdots\beta}-\cdots\nb\\
&+&
\Gamma^{\alpha}_{\mu\lambda}B^{i_1i_2\cdots\lambda}_{j_1j_2\cdots\beta} -
\Gamma^{\lambda}_{\mu\beta}B^{i_1i_2\cdots\alpha}_{j_1k\cdots\lambda},
\eqn
where $\Gamma^{\gamma}_{\mu\nu}$ is the space-time connection. Its form can be found by assuming that 
\bqn
\lb{Eq:CovMetZero}
D_{\alpha}g_{\mu\nu}=\nabla_{\alpha}g_{\mu\nu}=0. 
\eqn
Here we defined $\nabla$ to represent the space-time related component of the covariant derivative. The first equality is because the metric has no index of the tangent space.
Since we have imposed the torsion-free condition from the beginning, the connection would be the Christoffel symbol defined as
\bqn
\lb{Eq:DefChristoffel}
\Gamma^{\gamma}_{\mu\nu} \equiv \frac{1}{2}g^{\gamma\lambda}\left(
\partial_{\mu}g_{\lambda\nu}+\partial_{\nu}g_{\lambda\mu}-\partial_{\lambda}g_{\mu\nu}  
\right).
\eqn

The covariant derivative of the metric of the tangent space reads
\bqn
D_{\mu}\eta_{ij}&=&\partial_{\mu}\eta_{ij}-gA_{i~\mu}^{~k}\eta_{kj}-gA_{j~\mu}^{~k}\eta_{ik}\nb\\
&=&0,
\eqn
which is zero since $\partial_{\mu} \eta_{ij}=0$, and the Lorentz gauge field is anti-symmetric in the Latin indices.

We often impose an extra condition called the tetrad postulate that reads
\bqn
D_{\mu}e_{i\nu}&=&\partial_{\mu} e_{i\nu}-\Gamma^{\lambda}_{\mu\nu}e_{i\lambda}-
g A_{i~\mu}^{~k} e_{k\nu}\nb\\
&=&0.
\eqn
Note that we could have imposed this equation first and derive Eq.~\ref{Eq:CovMetZero}, but not vice versa. This tetrad postulate is an assumption about the equivalence of the coordinate and Lorentz frames and is imposed solely due to our physical intuition but not for mathematical consistency. Since the Christoffel symbols can be written entirely in terms of the metric, which itself is given in terms of the tetrad, the Lorentz gauge field and the tetrad were the two independent fields before imposing this condition. 

The strength tensor $F_{\mu\nu ij}$ is defined as
\bqn
\left[D_{\mu},D_{\nu}\right]\psi=\frac{g}{2}F_{\mu\nu ij}S^{ij}\psi,
\eqn
and is given by 
\bqn
\lb{eq:LorentzStrengthTensor}
F_{\mu\nu ij} = \partial_{\nu}A_{ij\mu}-\partial_{\mu}A_{ij\nu}-
gA_{i~\nu}^{~k}A_{kj\mu}+gA_{i~\mu}^{~k}A_{kj\nu}.\nb\\
\eqn
When expressed entirely using the coordinate indices, this is the Riemann curvature tensor
\bqn
R_{\mu\nu\alpha\beta} = e^i_{~\alpha}e^j_{~\beta}F_{\mu\nu ij}.
\eqn

Unlike in Yang-Mills theories, where the symmetry is uniquely associated with a single Lagrangian, in PGT, the symmetry allows more than one term. The most general form can be found in \cite{1995PhR...258....1H}
\bqn
\lb{LA}
{\cal{L}}_{A}=-\frac{1}{4}&\Big(&c_1 F_{\mu\nu ij}e^{i \mu}e^{j \nu}+c_2 F_{\mu\nu ij}F^{\mu \sigma ik}e^{j \nu}e_{k \sigma}\nb\\
&+&c_3 F_{\sigma \nu mj}F_{\mu \alpha in}e^{j \nu}e^{i \mu}e^{m \sigma}e^{n \alpha}\nb \\
&+&c_4F_{\mu \nu ij}F^{\alpha \beta mn}e^{i \mu}e^j_{~\beta}e_{m \alpha}e_n^{~\nu}\nb\\
&+&c_5F_{\mu \nu ij}F^{\mu\nu ij}\Big).
\eqn

If we had not imposed the torsion-free condition right from the beginning, the tetrad and the connection were still independent. Therefore, to find the field equations, we needed to vary the action with respect to both of them \cite{1995PhR...258....1H}. This is not the case anymore due to the torsion-free assumption and the tetrad postulate. Instead, the field equations should be derived by varying the action with respect to either the tetrad or the Lorentz gauge field. Since the tetrad postulate is a non-holonomic constraint, the results of the variation may depend on the variation path. This roots back to the fact that the non-holonomic constraint can be violated along an arbitrary displaced path \cite{greenwood2012classical,pars2013introduction,2005AmJPh..73..265F}.

The issue with the non-holonomic constraints can be resolved if the displacement paths in Hamilton's principle are restricted to those along which the constraint is not violated. 
To find such paths, we note that $D_{\mu} e_{i\nu} =0$ as the non-holonomic constraint is preserved under both coordinate and Lorentz frame transformations. Therefore, if we restrict the displacement paths to the ones generated by such transformations, the non-holonomic constraint will not be violated. 

The possible classes of displacement paths are 
\begin{enumerate}
\item\lb{Option1} option one: generated by coordinate transformations and leads to GR ,
\item\lb{Option2} option two: generated by homogeneous Lorentz transformation of the tetrad frames and leads to LGT \cite{2016CQGra..33b5008B,2016CQGra..33w5006B}.
\end{enumerate}

A more physically intuitive reason for the two options above is that while the torsion-free condition is imposed from the beginning, the framework still has two symmetries and, therefore, two independent conserved currents. In principle, each current can be used to generate a dynamic field. Since two dependent variables cannot have two independent sets of dynamical field equations, we have to choose one of the sources to drive the dynamics of the framework.

\section{Quantum LGT as a non-geometrodynamic approach to quantum gravity}
\label{sec:QLGT}
In the framework of option \ref{Option1}, the metric and its conjugate momenta are classically non-commutative, see the Poisson bracket in Eq.~\ref{Eq:GRPoisson}. Therefore, a quantum operator at microscopic scales should replace the metric, which means that the tetrad has quantum fluctuations and can be written as 
\bqn
e_{i\mu}=\e_{i\mu}+e^{\text{quantum}}_{i\mu}.
\eqn

In the framework of option \ref{Option2}, however, it is the Lorentz gauge field and not the tetrad that is classically non-commutative. Therefore, at the quantum level, it is the Lorentz gauge field and not the tetrad that should be replaced by a quantum operator. We now make two further postulates to build a self-consistent quantum theory of gravity.
\begin{itemize}
\item Postulate I: the tetrad has no quantum fluctuation even at the smallest length scales possible
\bqn
e_{i\mu}=\e_{i\mu}.
\eqn
\item Postulate II: the tetrad postulate is valid only at the classical level 
\bqn
&&D_{\mu}e_{i\nu} \neq 0,\lb{Eq:TetradPosInEquality}\\
&&\langle D_{\mu}e_{i\nu}\rangle=\partial_{\mu} e_{i\nu}-\Gamma^{\lambda}_{\mu\nu}e_{i\lambda}-
g\langle A_{i~\mu}^{~k} \rangle e_{k\nu}\lb{Eq:ClassicTetradPos}
=0.
\eqn  
\end{itemize}

We would like to mention that an assumption similar to our first postulate is the basis of the so-called quantum field theory in curved space-time. The difference is that in geometrodynamics theories, this assumption is valid only under specific semi-classical regimes but contradictory at the fundamental level due to the Poisson bracket in Eq.~\ref{Eq:GRPoisson}. In our non-geometrodynamics approach, the assumption can remain valid even at the fundamental level. This postulate alleviates the so-called problem of time because the time in our quantum theory of gravity has the same background nature as in quantum physics. 

The reason for the second postulate is that both the metric and the connection of space-time remain classical fields even at the quantum level due to the first postulate and Eqs.~\ref{Eq:MetricInTermOfTetrad}, and \ref{Eq:DefChristoffel}. On the other hand, the Lorentz gauge field of the tangent spaces is fundamentally a quantum operator.
Therefore, at the quantum level, the classical Christoffel symbols $\Gamma^{\alpha}_{\mu\nu}$ cannot be equivalent to the quantized Lorentz gauge field $A_{ij\mu}$ and the tetrad postulate cannot be valid.

In the Yang-Mills gauge theories based on the unitary groups, there is a unique possibility to form an invariant Lagrangian from a pair of strength tensors. The reason is that their group index is distinct from the space-time index. 
Space-time indices are coupled by the space-time metric while the group indices are coupled using the internal metric $\text{Tr}(\vec{t}\cdot\vec{t})$ with $\vec{t}$ being the generators of the group. The covariant derivatives of both the space-time and the internal metrics are zero. 

In LGT, in general, the Lagrangian is given in Eq.~\ref{LA} which has four independent invariant terms. The reason is that, in addition to the metric of space-time and the six-dimensional internal metric $\text{Tr}\left(S^{ij}\cdot S^{mn}\right)$, there exist the tetrad that can couple the gauge and space-time indices. 
However, at the quantum level, the covariant derivatives of the two metrics are zero, due to how the Christoffel symbol and the Lorentz gauge field are defined. But, the covariant derivative of the tetrad is not zero according to Eq.~\ref{Eq:TetradPosInEquality}, which provides a means of distinction between the terms. 
Therefore, to construct a Yang-Mills theory similar to those in the standard model of particle physics, we only allow those terms in the Lagrangian in which the strength tensors are coupled by objects whose covariant derivatives are zero. LGT is formally defined by 
\bqn
\lb{Eq:LGTLagrangian}
{\cal{L}}_{\text{A}} &=& \frac{1}{4}F_{\mu\nu ij}F_{~~~mn}^{\mu\nu}
\text{Tr}\left(S^{ij}\cdot S^{mn}\right)\nb\\
&=& \frac{1}{4}F_{\mu\nu ij}F^{\mu\nu ij},
\eqn 
where in the last line, we have used the six-dimensional internal metric of the homogeneous Lorentz group
\bqn
\lb{Eq:TraceOfLGT_Generators}
\delta^{ij,mn} \equiv \text{Tr}\left(S^{ij}\cdot S^{mn}\right)= 
\frac{1}{2}\left(\eta^{im}\eta^{jn}-\eta^{in}\eta^{jm}\right).
\eqn

In the Yang-Mills theories of the standard model, the sign of the Lagrangian is the opposite of what we introduced above and is to preserve the unitarity. Later in the paper, we discuss that since the Lorentz gauge fields are not tensors of the Lorentz group of physical observers and consequently cannot represent physical particles, the unitarity is preserved regardless of the sign in Eq.~\ref{Eq:LGTLagrangian}. Also, we will discuss that the definition is such that the classical field equations are in more agreement with GR.

\subsection{Path integral quantization of LGT}
\lb{sec:Quantization}
The total Lagrangian of QLGT reads
\bqn
\lb{Eq:TotalLagrangian}
&&{\cal{L}}_{\text{total}} =
\frac{i}{2}e_i^{~\mu}\bar{\psi}\gamma^i D_{\mu}\psi
-
\frac{i}{2}e_i^{~\mu}\bar{\psi}\overleftarrow{D}_{\mu}\gamma^i \psi
-
m\bar{\psi}\psi\nb\\
&&+
\frac{1}{4}F_{\mu\nu ij}F^{\mu\nu ij}
+
\frac{1}{2}\xi \left(\partial^{\mu}A_{ij\mu}\right)^2
-
\bar{c}^{ij}\partial^{\mu}\left(D_{\mu}c_{ij}\right),
\eqn
where in addition to the Lagrangian of LGT, we have added a gauge fixing terms with $\xi$ as an arbitrary parameter, and $c_{ij}$ with anti-symmetric indices represent the Faddeev-Popov ghosts. 
The generating functional, therefore, reads
\bqn
{\cal{Z}} = \int DA\, D\bar{\psi}\, D\psi \, D\bar{c}\, Dc\,
\exp\left(i\int e\, d^4x\, {\cal{L}}_{\text{total}} \right),
\eqn
where $e$ is the determinant of the tetrad. 
One crucial difference between this functional and the one in quantum-GR is that the path integration on the tetrad is absent since it is a background field at the fundamental level. In quantum field theories in curved space-time also $De_{i\mu}$ is absent. Nevertheless, that is only an approximation that is valid for particular situations and is contradictory at the fundamental level.

Since the metric is always a background field in QLGT, the quantization of gravity, i.e. $A_{ij\mu}$, in a non-flat background is similar to the quantization of other fields in curved space-time.

\subsection{QLGT in Minkowski space-time}
A study of quantum gravitational effects in the flat background is often considered essential \cite{Hooft2002} because it is easier to assess the underlying theory of gravity in this space-time. 
For instance, to renormalize GR the higher-derivative counter-terms that break the unitarity of the theory, and therefore are non-physical, are needed \cite{Hooft2002}. 
In the following, we evaluate the renormalizability and unitarity of QLGT in the flat background.

In the Minkowski space-time, $g_{\mu\nu} = \eta_{\mu\nu}$, $e=1$, $\Gamma^{\alpha}_{\mu\nu}=0$, and according to Eq.~\ref{Eq:ClassicTetradPos} $\langle A\rangle_{ij\mu} = 0$, just like the expectation value of the gauge fields of the standard model in the flat space-time.
The Feynman rules of QLGT are derived and discussed in appendix~\ref{app:FeynRules}, where we illustrate that the only notable difference between these rules and the ones in the standard model lies in their external lines. Since the gauge symmetry of QLGT is the Lorentz group of physical observers, its gauge fields cannot be associate with physical particles and do not receive external lines in the Feynman diagrams. 
Also, since the homogeneous Lorentz group is not compact, its internal metric is not positive definite, which implies that the corresponding gauge fields have negative kinetic energies. Nevertheless, the unitarity will not be violated since these gauge fields do not represent observable particles. 
In appendix~\ref{sec:unitarity}, the unitarity of QLGT is proved with a detailed calculation. 
Also, in appendix~\ref{sec:renormalizability}, we carry out a detailed calculation of the 1-loop divergencies of QLGT. We show that all of the infinities can be absorbed without a need for extra counter-terms. Therefore, QLGT is renormalizable up to the first loop level.


\section{The classical LGT}
\lb{sec:Classic}
In this section, we first eliminate the tetrad postulate to express the tetrad in terms of the dynamical field of LGT. Next, we derive the field equations, present a plane wave analysis, and enumerate a few exact solutions of LGT. 

\subsection{The tetrad in terms of the Lorentz gauge field}
In GR, the tetrad postulate is often used to eliminate the Lorentz gauge field in terms of the tetrad as the dynamical field. Similarly, in LGT, we need to eliminate the tetrad in terms of the Lorentz gauge field as the dynamical field.
We perform this elimination using the perturbation theory, which is enough for our purposes. 
We start with the assumption that the space-time fluctuations are small and expand the tetrad as 
\bqn
e_{i\mu} = \eta_{i\mu}+h_{i\mu}, 
\eqn
where $h \ll 1$, and we neglect the higher orders of $h$ in the following. Substituting this expression into Eqs. \ref{Eq:MetricInTermOfTetrad} and \ref{Eq:DefChristoffel}, the metric and the Christoffel symbols read
\bqn
&&g_{\mu\nu} = \eta_{\mu\nu} + 2 h_{(\mu\nu)},\nb\\
&&\Gamma^{\lambda}_{\mu\nu}=
\eta^{\lambda\sigma}\left(
\partial_{\mu}h_{(\sigma\nu)} + \partial_{\nu}h_{(\sigma\mu)} - 
\partial_{\sigma}h_{(\mu\nu)}
\right),
\eqn
where $h_{(\mu\nu)}\equiv\frac{1}{2}\left(h_{\mu\nu}+h_{\nu\mu}\right)$, and $h_{\mu\nu} \equiv \delta^i_{\mu}h_{i\nu}$. 
Substituting the tetrad, the metric, and the Christoffel symbols into the classical tetrad postulate in Eq.~\ref{Eq:ClassicTetradPos}, we can rewrite it as 
\bqn
\lb{Eq:LinearTetradPos}
&&\partial_{\mu}h_{i\nu} -\delta_i^{\sigma}\left(
\partial_{\mu}h_{(\sigma\nu)}+\partial_{\nu}h_{(\sigma\mu)}-\partial_{\sigma}h_{(\mu\nu)}
\right)
-g \A_{i\nu\mu}=0,
\eqn
where $\A_{i\nu\mu}\equiv\delta^j_{\nu}\A_{ij\mu}$. 

We separately multiply equation above by $\partial^i$ and $\partial^{\mu}$ to derive two equations, whose combination reads
\bqn
\Big(\eta^{\sigma}_i\partial^2-\partial_i \partial^{\sigma}\Big)h_{m\sigma} = 
g \partial^j\Big(\A_{j m i}-\A_{i m j}\Big).
\eqn 
The operator in the parentheses on the left hand side is the familiar operator from electromagnetism.
To uniquely determine $h_{m\sigma}$ as a solution to the equation above, we need to set a boundary condition. We note that this equation still refers to a local constraint rather than the dynamical field equation of a propagating field. Consequently, $h_{m\sigma}$ should be zero where $\A_{ij\mu}$ as its source is zero. Therefore, the corresponding Greens function $\Delta^{\gamma}_{\sigma}(y-x)\equiv \delta^{\gamma}_{\sigma} \Delta(y-x)$ should be equal to zero when $|y-x|>\delta$, where $\delta \simeq 0$ in comparison with macroscopic lengths.  
The tetrad uniquely reads
\bqn
\lb{Eq:TetradInTermsA}
e_{m\sigma}(x) &=& \eta_{m\sigma} 
+ g \int_{\delta} d^4y \partial_{_y}^j\Big( \A_{jm\sigma}(y)-\A_{\sigma mj}(y) \Big)
\Delta(y-x)
\nb\\
&\simeq & \eta_{m\sigma} 
+
g  \partial_{_x}^j\Big( \A_{jm\sigma}(x)-\A_{\sigma mj}(x) \Big) \int_{\delta} d^4y\Delta
.
\eqn
In the last line, we have assumed that the classical functions do not vary across the small distance $\delta\simeq 0$. From this equation, we can conclude that the space-time metric, the Christoffel symbols, and the Riemann tensor can all be expressed in terms of the Lorentz gauge field.

\subsection{Field equations}
The total action of LGT is
\bqn
I = \int e\, d^4x \left( {\cal{L}}_{A} + {\cal{L}}_{M} \right),
\eqn
where the Lagrangian of the gauge field is given in Eq.~\ref{Eq:LGTLagrangian}, and ${\cal{L}}_{M}$ is the Lagrangian of any regular matter field.
In the following, we first derive the linear field equations in a free-falling frame by dropping any term with $\A_{ij\mu}^3$ or higher from the total action and then carry out a coordinate and frame transformation to write the exact field equations. 

Since $F_{\mu\nu i j}$ is at least first order in $\A$, and the two Christoffel symbols in $F_{\mu\nu i j}$ have canceled each other out due to the index symmetries, to the quadratic order, we can write
\bqn
e\, {\cal{L}}_{A} \simeq \frac{1}{4}
\eta^{\mu\alpha}\eta^{\nu\beta}\eta^{i m}\eta^{j n}
\left( \partial_{\nu}\A_{ij\mu}-\partial_{\mu}\A_{ij\nu} \right)
\left( \partial_{\beta}\A_{mn\alpha}-\partial_{\alpha}\A_{mn\beta} \right).
\eqn
We perform a variation with respect to the dynamical field to get
\bqn
\delta I &=& \int d^4x \Bigg( 
\partial^{\nu} \bigg( \partial_{\nu}\A_{ij\mu}-\partial_{\mu}\A_{ij\nu} \bigg) \delta \A^{ij\mu}
 +
 \frac{\partial\left(\sqrt{-g} {\cal{L}}_{M}\right)}{\partial \A_{ij\mu}}
 \delta \A^{ij\mu}\nb\\
&-& g 
\partial^l \frac{\partial\left(\sqrt{-g} {\cal{L}}_{M}\right)}{\partial e_{m\alpha}}
 \bigg( \eta_{\big[ l i}\eta_{jm\big]}\eta_{\alpha\mu}-\eta_{\big[\alpha i}\eta_{jm\big]}\eta_{\mu l}\bigg) \delta \A^{ij\mu}\int_{\delta}d^4y\Delta
 \Bigg)\nb\\
&=&0,
\eqn
where in the last line we have used Eq.~\ref{Eq:TetradInTermsA} and performed an integration by parts. 
Therefore, the linear field equation reads
\bqn
\lb{Eq:linearFieldEq}
\partial^{\nu}F_{\mu\nu ij}&=&
-\frac{\partial\left(\sqrt{-g} {\cal{L}}_{M}\right)}{\partial \A_{ij\mu}}\nb\\
&-&  \frac{1}{2}g  \Big(\partial_j T_{i\mu} - \partial_i T_{j\mu}
+\partial_{\mu}\left(T_{ji}-T_{ij}\right)
\Big)\int_{\delta}d^4y\Delta,
\eqn
where the first term on the right hand side is the net spin of the classical matter and is zero unless the matter is artificially spin-polarized,  and $T^{m\alpha} \equiv \frac{\partial\left(\sqrt{-g} {\cal{L}}_{M}\right)}{\partial e_{m\alpha}}$ is conventionally defined as the energy-momentum tensor.

Using the equivalence principle, and the fact that the energy-momentum tensor is symmetric
in classical matter, we write the full equations as 
\bqn
\lb{Eq:EffectiveFieldEq0}
D^{\nu}F_{\mu\nu ij} = 
-\frac{\partial\left(\sqrt{-g} {\cal{L}}_{M}\right)}{\partial \A_{ij\mu}}
- \frac{1}{2}g
\Big(
D_j T_{\mu i} - D_i T_{\mu j}
\Big)\int_{\delta}d^4y\Delta.
\eqn
We can multiply equation above by $e_{~\alpha}^i e_{~\beta}^j$ to write it in the following convenient form
\bqn
\nabla^{\nu}R_{\mu\nu \alpha\beta} = 
-\frac{\partial\left(\sqrt{-g} {\cal{L}}_{M}\right)}{\partial \A_{\alpha\beta\mu}}
- \frac{1}{2}g
\Big(
\nabla_{\beta} T_{\mu \alpha} - \nabla_{\alpha} T_{\mu \beta}
\Big)\int_{\delta}d^4y\Delta.
\eqn
We would like to emphasize that this equation is not a higher derivative theory since, in LGT, the Lorentz gauge fields are the dynamical variables. 

At this point, we would like to make a connection between $\frac{1}{2}g \int_{\delta}d^4y\Delta$ and Newton's gravitational constant G. 
We multiply both sides of the equation above by $g^{\mu\alpha}$ and use the conservation law of the energy-momentum tensor to write
\bqn
\nabla^{\nu}R_{\nu\beta} = - \frac{1}{2}g  \nabla_{\beta}T^{\mu}_{\mu}\int_{\delta}d^4y\Delta,
\eqn
where $R_{\nu\beta}$ is the Ricci tensor and the spin source is dropped since it is zero when the matter is not spin-polarized. This equation is equal to the divergence of the Einstein equation in GR if we set 
\bqn
\frac{1}{2}g \int_{\delta}d^4y\Delta \equiv 4\pi G.
\eqn
Therefore, the classical field equation of LGT finally reads
\bqn
\lb{Eq:EffectiveFieldEq}
\nabla^{\nu}R_{\mu\nu \alpha\beta} = 
-\frac{\partial\left(\sqrt{-g} {\cal{L}}_{M}\right)}{\partial \A_{\alpha\beta\mu}}
- 4\pi G
\left(
\nabla_{\beta} T_{\mu \alpha} - \nabla_{\alpha} T_{\mu \beta}
\right).
\eqn

We now compare the equation above with the corresponding equation in GR. 
Contracting the second Bianchi identity, which holds for both GR and LGT, and using the Einstein equation in GR to eliminate the Ricci tensors, we can write
\bqn
\lb{Eq:GRBianchiEq}
\nabla^{\nu}R_{\mu \nu \alpha \beta} = 
-4\pi G \Big( 
\nabla_{\alpha}\left(2T_{\mu\beta}-Tg_{\mu\beta}\right)
-
\nabla_{\beta}\left(2T_{\mu\alpha}-Tg_{\mu\alpha}\right)
\Big),
\eqn
where $T\equiv T^{\mu}_{\mu}$.
Comparing the equation above with Eq.~\ref{Eq:EffectiveFieldEq}, we can conclude that they are the same when (i) the matter is not spin-polarized and (ii) the following equation is satisfied
\bqn
\lb{Eq:LGTGR_condition}
\nabla_{\beta}\left(T_{\mu\alpha}-\frac{1}{3}Tg_{\mu\alpha}\right)
\dot{=}
\nabla_{\alpha}\left(T_{\mu\beta}-\frac{1}{3}Tg_{\mu\beta}\right).
\eqn
The dot over the equality means that the two sides are not equivalent in general. One immediate conclusion is that when $T_{\mu\nu}=0$ or $T_{\mu\nu}\propto g_{\mu\nu}$, the Riemann tensor satisfies the same equations in GR and LGT.


It should be noted that the negative sign behind Newton's constant in Eq.~\ref{Eq:EffectiveFieldEq} is absent in \cite{2016CQGra..33w5006B}. The sign was originally chosen to be the same as in the gauge theories of the standard model where the unitarity condition requires it. 
However, as we discussed above, and also can be seen in appendix~\ref{sec:unitarity}, the sign of the gauge field Lagrangian of QLGT does not affect its unitarity.

\subsection{A plane wave analysis: helicity 2 space-time wave}
In geometrodynamic approaches to gravity, the observed space-time fluctuations are explained via the existence of massless gravitons just as the electromagnetic waves are explained via the existence of photons. In our framework, however, the Lorentz gauge fields do not represent physical particles. As a result, the framework has no candidate elementary particle for gravitation. Nevertheless, in this section, we show that the unobservable Lorentz gauge field drives helicity two waves of space-time that travel with the speed of light. 
We will show that LGT predicts the same propagating modes of space-time wave as in GR. The analogous plane wave analysis for GR can be found in \cite{1972gcpa.book.....W}.

We replace the fields in Eq.~\ref{Eq:LinearTetradPos} with the following Fourier transformations
\bqn
&&h_{i\mu}=\int \frac{d^4p}{(2\pi)^4}e^{-ip\cdot x}\Sigma_{i\mu}(p),\\
&&\A_{ij\mu} = \int \frac{d^4p}{(2\pi)^4}e^{-ip\cdot x}\varepsilon_{ij\mu}(p)\lb{Eq:GaugeFieldExpansion}.
\eqn
After multiplying by $e^{ik\cdot x}$ and integrating on the position space, the linear form of the classical tetrad postulate in the momentum space reads
\bqn
\lb{Eq:tetradPosInMomentumSpace}
ik_{\mu}\Sigma_{[\nu i]}+ik_{\nu}\Sigma_{(i\mu)}-ik_i \Sigma_{(\mu\nu)}=g\varepsilon_{i\nu\mu},
\eqn
where the brackets around the indices indicate anti-symmetrization, and both $\Sigma$ and $\varepsilon$ are functions of the same momentum $k$.

Using Eq.~\ref{Eq:linearFieldEq} and Eq.~\ref{Eq:GaugeFieldExpansion} and after fixing the Lorentz gauge by $\partial^{\mu}\A_{ij\mu} = 0$, 
we can conclude that the wave travels at the speed of light $k^2=0$, and $k^{\mu}\varepsilon_{ij\mu}=0$. 
Without loss of generality, we assume that the spatial component of the momentum vector is in the z-direction. Therefore, we can conclude that 
\bqn
&&k^{\mu}=(k,0,0,k),\lb{Eq:MasslessMomentum}\\
&&\varepsilon_{ij0}=-\varepsilon_{ij3}\lb{Eq:TransverseWave}.
\eqn  
Also, we fix the space-time symmetry by setting $g^{\mu\nu}\Gamma^{\lambda}_{\mu\nu}=0$ to write
\bqn
\lb{Eq:PlaneWaveSpaceFix}
2k^{\mu}\Sigma_{(\mu\sigma)}=k_{\sigma}\eta^{\mu\nu}\Sigma_{\mu\nu}.
\eqn

The physical modes of space-time fluctuations cannot be eliminated if we wish to make a second coordinate or homogeneous Lorentz transformation, which in the momentum space are 
\bqn
\lb{Eq:MomentumSpaceTransformations}
&&\Sigma'_{i\mu} \rightarrow \Sigma_{i\mu} - ik_{\mu}\xi_i(k),\nb\\
&&\varepsilon'_{ij\mu} \rightarrow \varepsilon_{ij\mu} + ik_{\mu}\omega_{ij}(k),
\eqn
where the four components of $\xi$ and the six components of $\omega$ are arbitrary. Since $k^{\mu}=(k,0,0,k)$, the equation above indicates that any of the components of $\Sigma_{i\mu}$ and $\varepsilon_{ij\mu}$ with $\mu$ equal to 0 or 3 can be eliminated by an appropriate choice of the free parameters. 
Therefore, the physical components of Eqs.~\ref{Eq:tetradPosInMomentumSpace},~\ref{Eq:MasslessMomentum}, and \ref{Eq:PlaneWaveSpaceFix}, 
or equivalently their un-compacted version in Eq.~\ref{Eq:SolutionToTetPos}, read
\bqn
\lb{Eq:PhysicalModes}
&&\Sigma_{11}=\frac{i}{k}\varepsilon_{011},\nb\\
&&\Sigma_{22}=\frac{i}{k}\varepsilon_{022},\nb\\
&&\Sigma_{12}+\Sigma_{21}=\frac{2i}{k}\varepsilon_{021}.
\eqn
Note that these are the symmetrized components of the tetrad equivalent to three of the components of the metric. It is interesting to note that these are the physical components of the gravitational wave in GR as well.

Using Eq.~\ref{Eq:PlaneWaveSpaceFix}, or equivalently Eq.~\ref{Eq:SpaceTimeGaugeFixing}, we can eliminate $\Sigma_{22}$ such that only $g_{11}$ and $g_{12}$ are the independent physical modes of the metric. We define the following two fields in terms of the two physical components of the metric
\bqn
\Sigma_{\pm} \equiv \Sigma_{11} \mp i \Sigma_{(12)}. 
\eqn
By performing a rotation around the direction of the motion of the wave by an angle $\theta$, we can show that 
\bqn
\Sigma'_{\pm} \rightarrow e^{\pm 2i\theta}\Sigma_{\pm},
\eqn
which means that the physical modes of the metric have helicity two. 

Therefore, by using the field equation of LGT together with the tetrad postulate, we have concluded that space-time plane waves have helicity-2, the result that is often derived using the Einstein field equations.

\subsection{A few exact solutions of LGT}
In the following, we mention five space-time metrics and discuss the necessary conditions for them to be exact solutions of LGT. 
Calculations are all carried out using computer packages, and the corresponding scripts are available in \cite{ComputerScriptsBorzou}.

\paragraph{The de Sitter space-time:}
The metric of the de Sitter space is
\bqn
ds^2 = -dt^2+e^{Ht}\big|\vec{dx}\big|^2,
\eqn
where $H$ is a constant. After substituting this metric into Eq.~\ref{Eq:EffectiveFieldEq}, and dropping the spin term, we observe that the source needs to be zero for the metric to be a solution. On the other hand, even in the presence of the vacuum energy, the source is zero \cite{2017CQGra..34n5005B}. Therefore, unlike in GR, we not only need no dark energy to explain an accelerating expansion of the universe but also the prediction of the standard model of particle physics for the magnitude of the vacuum energy is not contradictory anymore.

\paragraph{The early universe solution:}
The so called radiation dominated universe in $\text{GR}-\Lambda-\text{CDM}$ model has the following form
\bqn
ds^2 = -dt^2 + b\cdot t \big|\vec{dx}\big|^2,
\eqn
where $b$ is a constant. We have substituted this metric into Eq.~\ref{Eq:EffectiveFieldEq} and have shown that this is an exact solution if the source is zero. Moreover, even in the presence of radiation the source remains zero \cite{2017CQGra..34n5005B}, which means that, unlike in GR, the early universe solution in LGT stays stable even if other light particles exist or if the neutrinos are not hot.

\paragraph{The matter-dominated universe:}
The metric in the matter-dominated universe reads
\bqn
ds^2 = -dt^2 + b\cdot t^{\frac{4}{3}} \big|\vec{dx}\big|^2.
\eqn
The needed energy-momentum tensor for this solution is the same in both LGT and GR.

\paragraph{The Schwarzschild solution in vacuum:}
The metric for this space-time is 
\bqn
ds^2 &=& -\left(1-\frac{2\text{GM}}{r}\right)dt^2
+\left(1-\frac{2\text{GM}}{r}\right)^{-1}dr^2
+r^2d\Omega^2,
\eqn
which is an exact solution in both GR and LGT with zero energy-momentum tensors. This solution is previously discussed in \cite{2016CQGra..33b5008B}.

\paragraph{The Schwarzschild solution inside a star:}
The metric for a spherically symmetric solution inside an incompressible star reads
\bqn
ds^2 &=&
-\left(\frac{3}{2}\sqrt{1 - \frac{2GM}{R}} - \frac{1}{2}\sqrt{1 - \frac{2GM}{R^3}r^2}\right)^2 dt^2 \nb\\
&+&
\frac{1}{1-\frac{2GM}{R^3}r^2}dr^2+ r^2 d\Omega^2.
\eqn
In both GR and LGT, this is an exact solution with the same energy-momentum tensor that reads 
$T^{\mu}_{\nu}=\left(-\rho,p,p,p\right)$, where $\rho$ is constant and 
\bqn
p = \rho\Bigg( \frac{\sqrt{1 - \frac{2GM}{R}} - \sqrt{1 - \frac{2GM}{R^3}r^2}}
{\sqrt{1 - \frac{2GM}{R^3}r^2}-3\sqrt{1 - \frac{2GM}{R}}} \Bigg).
\eqn

\paragraph{The Kerr metric:}
The Kerr space-time is crucial for describing some astrophysical observations, and its line element reads
\bqn
ds^2 &=& 
-\left(1-\frac{2GMr}{r^2+a^2\cos^2\theta}\right)dt^2
-\frac{4GMar\sin^2\theta}{r^2+a^2\cos^2\theta}dtd\phi\nb\\
&+&\frac{r^2+a^2\cos^2\theta}{r^2-2GMr+a^2}dr^2+\left(r^2+a^2\cos^2\theta\right)d\theta^2\nb\\
&+&\Big(
 \left(a^4+a^2r^2\right)\cos^2\theta + 2GMra^2 \sin^2\theta+a^2r^2+r^4
\Big)\nb\\
&&\times \frac{\sin^2\theta}{r^2+a^2\cos^2\theta} d\phi^2.
\eqn
This also is an exact solution of both LGT \cite{KerrMarzieMehrdad} and GR with zero energy-momentum tensors.

At the end of this section, we would like to mention that the classical solutions of the field equations of PGT have been studied in several places. See \cite{1980grg1.conf..329H,1989GReGr..21.1107O,2017PhRvD..96f4031B} and the references therein for an incomplete list. 
However, the field equations of LGT are different from those of PGT for a few reasons. First, the Lagrangian of LGT is only one of the terms in PGT Lagrangian. The second reason is our different choice of the dynamical variables and the complexities of the non-holonomic constraint.

\section{Experimental tests of LGT}
\lb{Sec:Experiment}
LGT passes all of the gravitational tests that GR has passed so far because (i) both the Schwarzschild and the Kerr space-times are its exact solutions, (ii) it can explain the direct observation of space-time waves. In this section, we discuss possible experiments that can differentiate between the two theories.

Both GR and LGT describe space-time in terms of a pseudo-Riemannian geometry where the geodesic deviation satisfies the following equation 
\bqn
\frac{d^2\xi^{\mu}}{d\tau^2} = 
- R^{\mu}_{~\alpha \beta \gamma}\frac{dx^{\alpha}}{d\tau}\frac{dx^{\gamma}}{d\tau}\xi^{\beta},
\eqn
where $x^{\mu}(\tau)$ and $x^{\mu}(\tau)+\xi^{\mu}(\tau)$ are the geodesics of two nearby particles. 
On the other hand, Eqs.~\ref{Eq:EffectiveFieldEq} and \ref{Eq:GRBianchiEq} indicate that in the presence of a spin-polarized matter, or when Eq.~\ref{Eq:LGTGR_condition} does not hold, the Riemann tensors in LGT and GR have different values. Hence, observing geodesic deviations under the mentioned conditions can distinguish between the two theories.

In LGT, both the spin and the mass of particles are the sources of gravity. Since one of the sources in Eq.~\ref{Eq:EffectiveFieldEq} is fully anti-symmetric while the other is partially symmetric, there is no spin-mass force in LGT. However, the spin-spin force of LGT can be tested. 
The corresponding Feynman vertex is given in Eq.~\ref{Eq:VAFF}, which indicates that the spin-spin force, in the static situation, is similar to the Coulomb force with the Coulomb constant replaced by $\frac{g}{4}$, and the electric charges replaced by the net number of the spin of matters. Even though $g$ needs to be small enough to satisfy the current accelerator-based constraints, we can generate an observable force by increasing the net number of spins in the matter by artificially polarizing it. 
To reduce the background from the short-range electromagnetic spin-spin interactions, the polarized matters should be spatially separated. The standard model background can be further reduced if the experiment is performed on low energy neutrinos by passing them from the vicinity of, but not through, a highly spin-polarized matter. 
The observable effects can be magnified if the neutrino beam travels a long distance after it is deflected by the spin-spin force. 
This experiment can be performed with the current technology since many underground neutrino experiments around the world send neutrino beams from a city and detect it in another city. Also, many labs around the world can spin-polarize matter.

It is also possible to use the cosmological and astronomical observations to differentiate between GR and LGT.
As we discussed earlier in this paper, the expansion of the universe depends on its radiation content in GR but is independent of its radiation content in LGT \cite{2018EPJC...78..639B}. 
On the other hand, the observation of the light elements left from the Big Bang nucleosynthesis (BBN) is highly sensitive to the expansion profile of the universe.
Therefore, a light dark matter can put GR-based cosmology in conflict with the BBN observations. Resolving such a conflict is harder than resolving the current issues in cosmology because subtracting from the energy content of the universe is harder than adding to it. 
Recently, we have placed an upper bound on the mass of dark matter by arguing that its temperature in galactic halos cannot be negative \cite{2020arXiv200304532B}.
Using the observations of the Milky Way, the upper bound on the mass of dark matter turns out to be 
\bqn
\lb{Eq:DMUpperMass}
m < 542 (\text{eV}) \times T_0 (\text{Kelvin}),
\eqn
where $T_0$ is the temperature of dark matter at the center of the Milky Way. 
If the cold dark matter had not been trapped in galaxies, its current temperature was nearly zero Kelvin. Since dark matter is blind to the standard model forces, it can be heated up only through the gravitational-based mechanisms.   
Although a conclusion cannot be drawn yet, $T_0$ of the Milky Way may not even exceed a few Kelvins given that the Newtonian gravity is extremely weak. This analysis awaits an estimation of the temperature profile of galactic halos. If $T_0$ turns out to be small, dark matter is light and GR-based cosmology will conflict with the BBN observations while LGT-based cosmology will not \cite{2018EPJC...78..639B}.

\section{Conclusion}
\lb{sec:conclusion}
We have presented QLGT, a quantum Yang-Mills theory of gravity based on the homogeneous Lorentz group of the physical observers in the tangent spaces of a torsion-free pseudo-Riemannian manifold. Therefore, the dynamics of the theory is defined on space-time but distinct from it. In the macroscopic world, the dynamics of the tangent spaces is coupled to space-time through the tetrad postulate as a non-holonomic constraint. Therefore, unlike the conventional Poincare gauge theory approaches, the space-time metric does not participate in any Poisson bracket in the classical regimes. This allows us to postulate, without contradicting the quantum principles, that space-time has no quantum nature at all, even in the highest possible energies. 
Therefore, the nature of time in QLGT is the same as in quantum physics, and the long-stood problem of time is alleviated.

We have derived the Feynman rules of QLGT and calculated the one-loop diagrams. We have shown that QLGT is renormalizable to the first loop approximation. The unitarity of the theory has been studied. We have shown that the probabilities are conserved in QLGT. These are non-trivial results since the homogeneous Lorentz group is non-compact, and its internal six-dimensional metric is not positive-definite. The difference between our Yang-Mills theory and a typical Yang-Mills theory of the homogeneous Lorentz group is that we are utilizing the group of physical observers. Consequently, our Lorentz gauge fields do not represent elementary particles, and no particle with negative kinetic energy exists in our scenario.  

The classical field equations of LGT have been presented as well. Using a plane wave analysis, we have shown that the dynamics of non-physical Lorentz gauge fields in the tangent spaces are transferred to space-time using the equation of tetrad postulate. As a result, while gravity is associated with no elementary particle, space-time waves are generated by the fluctuations of an unobservable gravitational field that lives on space-time. We have shown that the physical modes of space-time wave in LGT are the same as the physical modes of space-time wave in GR, and their helicities are equal to two. 

Also, we have shown that the classical field equations of LGT possess the Schwarzschild solutions in the vacuum and inside the stars, the Kerr solution, the de Sitter solution, and the matter-dominated and the early universe space-time solutions in their exact forms.
Although LGT and GR share many exact solutions, it is still possible to differentiate between them experimentally. We have discussed a few such experiments in the present paper.

\appendix

\section{An overview of quantization of GR}
\renewcommand{\theequation}{A.\arabic{equation}} \setcounter{equation}{0}
\lb{sec:QuantizationOfGeometry}
In the last part of this section, we would like to review the quantization of option \ref{Option1} above. GR is uniquely the simplest theory of a massless spin two elementary particle \cite{1954PhRv...96.1683G,1955PhRv...98.1118K,1965PhRv..138..988W,1970GReGr...1....9D,1975AnPhy..89..193B,doi:10.1063/1.524007} where only $c_1$ is kept non-zero in Eq.~\ref{LA}. For the quantization purposes, it is easier to decompose the Lagrangian as in \cite{1962rdgr.book..127A}
\bqn
&&S_{\text{GR}}=\int {\cal{L}}_{\text{GR}} dtd^3x,\nb\\
&&{\cal{L}}_{\text{GR}}=\left(R^{(3)}+K_{ab}K^{ab}-K^a_aK^b_b\right)N \sqrt{g^{(3)}},
\eqn
where $R^{(3)}$ refers to the three-dimensional curvature, $K_{ab}$ is the extrinsic curvature, $g^{(3)}$ is the determinant of $g^{(3)}_{ab}$, and the rest of the parameters are defined in the following decomposition of the space-time metric
\bqn
ds^2 = -\left(Ndt\right)^2
+
g^{(3)}_{ab}\left(dx^a+N^adt\right)\left(dx^b+N^bdt\right).
\eqn
The dynamical variables are $N$, $N^a$, and $g^{(3)}_{ab}$. The corresponding canonical momenta are 
\bqn
\pi = \frac{\partial {\cal{L}}_{\text{GR}}}{\partial \dot{N}},~~
\pi_a = \frac{\partial {\cal{L}}_{\text{GR}}}{\partial \dot{N}^a},~~
\pi^{ab} = \frac{\partial {\cal{L}}_{\text{GR}}}{\partial \dot{g}^{(3)}_{ab}}.
\eqn
The first two momenta are zero and define the constraints of GR. 
This is similar to the case of electrodynamics, where the canonical momentum of the temporal component of the vector potential is zero.
The following Poisson bracket drives the dynamics of GR \cite{kiefer2007quantum}
\bqn
\lb{Eq:GRPoisson}
\Big\{g^{(3)}_{ab}\left(\vec{x}\right),\pi^{ij}\left(\vec{y}\right)\Big\}
= \frac{1}{2}\left(\delta^i_a\delta^j_b+\delta^i_b\delta^j_a\right)
\delta^3\left(\vec{x}-\vec{y}\right). 
\eqn

To quantize the theory, we replace the dynamical variables by the corresponding quantum operators and the Poisson bracket by the canonical commutation relation. 
In the presence of the constraints mentioned above, such canonical quantization is cumbersome but informative and can be found in \cite{1987quco.book...93D}. With the developments in the standard model of particle physics, and especially after the works of Feynman \cite{Feynman:1963ax}, Faddeev and Popov \cite{1967PhLB...25...29F}, Mandelstam  \cite{1968PhRv..175.1604M}, and DeWitt \cite{1968PhRv..171R1834D}, the path integral formalism, also called the manifestly covariant method, became the standard method of carrying out this quantization. 

In the path integral quantization, the four dimensional space-time metric as the dynamical field, is first expanded around a background, which for simplicity we take to be flat,
\bqn
g_{\mu\nu}=\eta_{\mu\nu}+ f_{\mu\nu},
\eqn
where $\eta$ is the Minkowski metric. The Lagrangian of GR is rewritten in terms of $f_{\mu\nu}$ and the generating functional reads
\bqn
{\cal{Z}}\left[t^{\mu\nu}\right] = \int {\cal{D}}f_{\mu\nu}
\exp\left(
i\int d^4x \left[{\cal{L}}_{\text{GR}}+f_{\mu\nu}t^{\mu\nu}\right]
\right). 
\eqn
Due to the diffeomorphism invariance of the theory, we also need to fix the gauge. A common choice is to use the harmonic coordinates where 
\bqn
C_{\nu}\equiv \partial^{\mu}f_{\mu\nu}-\frac{1}{2}\partial_{\nu}f^{\mu}_{\mu}=0.
\eqn
Therefore, the generating functional receives a corresponding gauge fixing and a Faddeev-Popov ghost Lagrangian
\bqn
{\cal{L}}_{\text{GR}} \rightarrow 
{\cal{L}}_{\text{effective}} \equiv
{\cal{L}}_{\text{GR}}
+{\cal{L}}_{\text{GF}}
+{\cal{L}}_{\text{FP}}.
\eqn
The Feynman rules are subsequently read from the effective Lagrangian. These sets of rules can then be used to calculate the Feynman diagrams that contain loops. Some of the loop diagrams are divergent, as in other field theories. However, in \cite{1974AIHPA..20...69T,1986NuPhB.266..709G}, the authors have shown that the infinities cannot be removed by adding counterterms that have the same form as in the effective Lagrangian. Therefore, GR is a non-renormalizable theory with no falsifiable prediction for high energies. Such ultraviolet divergencies are expected in any theory like GR, whose coupling constant has a negative dimension in the mass units \cite{1979grec.conf..790W}. 

Even if GR was a renormalizable field theory, we still did not have a consistent quantum theory of gravity. Because, in such quantum gravity, on the one hand, time is a classical background, and on the other hand, it is a quantum operator. The problem of time in quantum geometrodynamics is extensively studied but is still open. A review of the subject can be found in \cite{kiefer2007quantum,1992gr.qc....10011I,rovelli2007quantum}. 
Moreover, unlike particle physics' standard model, GR is driven by the absolute value of energies, instead of their differences. This has led to the cosmological constant problem  \cite{1989RvMP...61....1W}. 

In the end, we refer the reader to \cite{2002PhR...357..113S,2014CQGra..31r5002S,2018PhRvD..98b4014B,2019PhRvD..99f4001L}, and the references therein, for an overview of the quantum aspects of the broader Poincare gauge theories.

\section{Feynman rules of QLGT}
 \renewcommand{\theequation}{B.\arabic{equation}}
 \setcounter{equation}{0}
\lb{app:FeynRules}
 
\paragraph{Propagators\\}  
The inverse propagator of the Lorentz gauge field in the momentum space reads
\bqn
\left(\text{PA}(k)^{-1}\right)^{ij\mu,mn\nu}=\frac{\delta^2 {\cal{L}}_{\text{total}}}{\delta A_{ij\mu}(k)\delta A_{mn\nu}(k)}\Big\vert_{A=\psi=c=0},
\eqn
where $\textsf{PA}$ stands for the propagator for field $A_{ij\mu}$. 
We use the FeynCalc package \cite{1991CoPhC..64..345M,2016CoPhC.207..432S} to take the two functional derivatives, and have made the scripts available online \cite{ComputerScriptsBorzou}. The propagator reads
\bqn
\textsf{PA}(k)_{ij\mu,mn\nu} &=& i\delta_{ij,mn}
\left(\frac{\eta_{\mu\nu}}{k^2}-(1-\xi)\frac{k_{\mu}k_{\nu}}{k^4}\right).
\eqn
The propagator for the Faddeev-Popov ghosts can be found via the same procedure
\bqn
\textsf{Pc}(p)_{ij,mn}=\frac{i\delta_{ij,mn}}{p^2}.
\eqn
The fermion propagator is also known to be 
\bqn
\textsf{PF}(p)=i\frac{p\cdot\gamma+m}{p^2-m^2}.
\eqn

\paragraph{Vertices\\}
\lb{Sec:Vertices}
The interactions in QLGT can be found via higher-order Functional derivatives. 
Due to their somewhat lengthy nature, we calculate them with the FeynCalc package again and make them available in \cite{ComputerScriptsBorzou}. 

The interaction with fermions read
\bqn
\lb{Eq:VAFF}
&&\text{VAFF}^{ij\mu}\equiv
\frac{i\delta^3{\cal{L}}_{\text{total}}}{\delta \bar{\psi}\delta A_{ij\mu}\delta \psi}\Big\vert_{A=\psi=c=0}=\nb\\
&&\nb\\
&&~~~~~~~~~\begin{fmffile}{AFFVertex}
      \setlength{\unitlength}{0.55cm}
      \parbox{30mm}{\begin{fmfgraph*}(3.5,3.5)
  \fmfright{i1,i2}
  \fmfleft{o}
  \fmf{fermion}{i1,v}
  \fmf{wiggly}{o,v}
  \fmf{fermion}{v,i2}
        \fmflabel{$ij\mu$}{o}
      \end{fmfgraph*}}
    \end{fmffile}.
\eqn
The naming VAFF stands for the vertex of a gauge field and two fermions. 
The interaction of the Faddeev-Popov ghosts with the gauge field is
\bqn
&&\text{VAc}\bar{\text{c}}(p)^{a_1 b_1, a_2 b_2 \mu_2, a_3 b_3}\equiv\nb\\
&&\frac{i\delta^3{\cal{L}}_{\text{total}}}{\delta \bar{c}(p)_{a_1b_1}\delta A_{a_2b_2\mu_2}\delta c_{a_3b_3}}\Big\vert_{A=\psi=c=0}=\nb\\
&&~~\nb\\
&&~~\nb\\
&&~~~~~~~~~~~~~
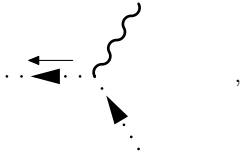
\begin{fmffile}{GhostVertex}
      \setlength{\unitlength}{0.55cm}
      \parbox{30mm}{\begin{fmfgraph*}(3.5,3.5)
  \fmfright{i1,i2}
  \fmfleft{o}
  \fmf{ghost}{i1,v}
  \fmf{ghost,label=$p$}{v,o}
  \fmf{wiggly}{i2,v}
        \fmflabel{$a_3b_3$}{i1}
        \fmflabel{$a_2b_2\mu_2$}{i2}
        \fmflabel{$a_1b_1$}{o}
        \marrow{c}{up}{top}{}{v,o}
      \end{fmfgraph*}}
    \end{fmffile},\nb\\
&&~~\nb\\
\eqn
The following is the gauge field interaction of order $g$
\bqn
&&\text{V3A}(k_1,k_2,k_3)^{i_1j_1\mu_1,i_2j_2\mu_2,i_3j_3\mu_3} \equiv\nb\\
&&\frac{i\delta^3{\cal{L}}_{\text{total}}}{\delta A_{i_1j_1\mu_1}(k_1)\delta A_{i_2j_2\mu_2}(k_2)\delta A_{i_3j_3\mu_3}(k_3)}\Big\vert_{A=\psi=c=0}=\nb\\
&&\nb\\
&&~\nb\\
&&~\nb\\
&&~~~~~~~~
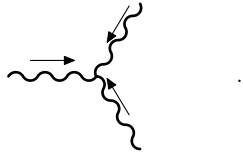
\begin{fmffile}{A3Vertex}
      \setlength{\unitlength}{0.55cm}
      \parbox{30mm}{\begin{fmfgraph*}(3.5,3.5)
  \fmfright{i1,i2}
  \fmfleft{o}
  \fmf{wiggly,label=$k_3$,label.side=right}{i1,v}
  \fmf{wiggly,label=$k_1$,label.side=right}{o,v}
  \fmf{wiggly,label=$k_2$}{i2,v}
  \marrow{a}{up}{top}{}{i1,v}
  \marrow{b}{up}{bot}{}{o,v}
  \marrow{c}{up}{top}{}{i2,v}
        \fmflabel{$i_3j_3\mu_3$}{i1}
        \fmflabel{$i_2j_2\mu_2$}{i2}
        \fmflabel{$i_1j_1\mu_1$}{o}
      \end{fmfgraph*}}
    \end{fmffile}.\nb\\
    &&~
\eqn
Finally, gauge field interaction of order $g^2$ is equal to
\bqn
&&\text{V4A}^{i_1j_1\mu_1,i_2j_2\mu_2,i_3j_3\mu_3,i_4j_4\mu_4} \equiv\nb\\
&&\frac{i\delta^4{\cal{L}}_{\text{total}}}{\delta A_{i_1j_1\mu_1}\delta A_{i_2j_2\mu_2}\delta A_{i_3j_3\mu_3}\delta A_{i_4j_4\mu_4}}\Big\vert_{A=\psi=c=0}=\nb\\
&&~\nb\\
&&~\nb\\
&&~~~~~~
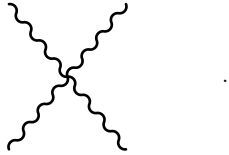
\begin{fmffile}{A4Vertex}
      \setlength{\unitlength}{0.55cm}
      \parbox{30mm}{\begin{fmfgraph*}(3.5,3.5)
  \fmfright{i1,i2}
  \fmfleft{o1,o2}
  \fmf{wiggly}{i1,v}
  \fmf{wiggly}{o1,v}
  \fmf{wiggly}{o2,v}
  \fmf{wiggly}{i2,v}
        \fmflabel{$i_3j_3\mu_3$}{i1}
        \fmflabel{$i_2j_2\mu_2$}{i2}
        \fmflabel{$i_4j_4\mu_4$}{o1}
        \fmflabel{$i_1j_1\mu_1$}{o2}
      \end{fmfgraph*}}
    \end{fmffile}.\nb\\
&&~\nb\\
\eqn

\paragraph{External lines\\}
\lb{subsec:ExternalLine}
The total Lagrangian in flat space-time is a function of the Faddeev-Popov ghosts, the fermions, and the Lorentz gauge field.
The Faddeev-Popov ghosts have wrong statistics and cannot represent physical particles. Therefore, they do not receive an external line in the Feynman diagrams. Also, following the convention, we show incoming and outgoing fermions with $u^{\sigma}(p)$ and $\bar{u}^{\sigma}(p)$ respectively, while incoming and outgoing anti-fermions with $\bar{v}^{\sigma}(p)$ and $v^{\sigma}(p)$, where $\sigma$ refers to the two spin modes.

To discuss the external lines for the Lorentz gauge field, we note that if the arbitrary parameter of the Lorentz transformation of physical observers $\omega_{ij}$ is much smaller than unity, Eq.~\ref{Eq:GaugeFieldGeneralTransform} implies that the Lorentz gauge field transforms as 
\bqn
\tilde{A}_{ij\mu}(x) = A_{ij\mu}(x) + D_{\mu}\omega_{ij}, 
\eqn
which has an identical form as the transformation of the gauge fields in the standard model under a special unitary gauge transformation with parameter $\alpha^a$
\bqn
\grave{A}^a_{\mu}(x) = A^a_{\mu}(x) + D_{\mu}\alpha^a.
\eqn
However, despite the similarity, there is a crucial difference. 
Under a Lorentz transformation of physical observers, $\Lambda^i_{~j} \simeq \delta^i_j+\omega^i_{~j}$, the gauge field of the unitary group has an invariant length equal to
\bqn
\left(A^a_{\mu}A^{a\mu} + 2D_{\mu}\alpha^a A^{a\mu} + D_{\mu}\alpha^a D^{\mu}\alpha^a\right)^{\frac{1}{2}},
\eqn
which is independent of $\omega^i_{~j}$. This means that two independent experiments observe the same length for $\grave{A}^a_{\mu}$.
On the other hand, the length of $\tilde{A}_{ij\mu}$ is equal to 
\bqn
\left(A_{ij\mu}A^{ij\mu} + 2D_{\mu}\omega_{ij} A^{ij\mu} + D_{\mu}\omega_{ij} D^{\mu}\omega^{ij}\right)^{\frac{1}{2}},
\eqn
which depends on the parameter of the Lorentz transformation of physical observers. Therefore, the Lorentz gauge field is observer-dependent, does not have an invariant length and, consequently, cannot represent physical particles. Therefore, it receives no external line in the Feynman diagrams of QLGT. Later in this paper, we discuss that the Lorentz gauge field induces classical helicity two space-time fluctuations, which are observable fields.

\section{Unitarity of QLGT}
 \renewcommand{\theequation}{C.\arabic{equation}} \setcounter{equation}{0}
\label{sec:unitarity}

To preserve probability in a quantum field theory, the S matrix has to be unitary. This means that
\bqn
2 \text{Im}\left(T\right) = TT^{\dagger},
\eqn
where $T\equiv -i(\text{S}-1)$. 
If $|\phi \rangle$ is a state of the system, the equation implies that
\bqn
\lb{Eq:OpticalTheorem}
2 \text{Im}\langle\phi| T|\phi\rangle = \sum_k\langle \phi| T|k\rangle\langle k|T^{\dagger}|\phi\rangle,
\eqn
where $|k\rangle$ refers to the physically observable modes of the fields and $\sum_k |k\rangle\langle k|=1$. In the gauge theories of the standard model, only fermions and the transverse component of the gauge fields contribute to the $|k\rangle$ states while the non-physical longitudinal components of the gauge fields and Faddeev-Popov ghosts are excluded. 

In QLGT, the Lorentz gauge field cannot represent a physical mode and should be excluded entirely from the states of $|k\rangle$. Therefore, the only physical states are the fermions, and consequently, Eq.~\ref{Eq:OpticalTheorem} is non-trivial only if $1=\sum_k |k\rangle\langle k|$ is intervening internal fermionic lines of Feynman diagrams. 
We now prove the unitarity for a rather general Feynman diagram of the form
\bqn
&&~~~~~~~~~\begin{fmffile}{UnitarityFig}
      \setlength{\unitlength}{0.55cm}
      \parbox{30mm}{\begin{fmfgraph*}(3.5,3.5)
\fmfleft{i} 
\fmfright{o} 
\fmf{wiggly,label=$k$}{i,v1} 
\fmf{wiggly,label=$k$}{v2,o}
\fmf{fermion,left,tension=.3}{v1,v2,v1}  
\fmflabel{?}{i}   
\fmflabel{?}{o} 
\end{fmfgraph*}}
\end{fmffile},
\eqn
where the question marks can be any multi-loop diagram and in the following will be presented by ${\cal{M}}_{1_{ij\mu}}$. The amplitude for diagram above reads
\bqn
&&i{\cal{M}}=-\int \frac{d^4q}{(2\pi)^4} {\cal{M}}_{1_{i_1j_1\mu_1}}{\cal{M}}^{\ast}_{1_{i_2j_2\mu_2}}\text{Tr}\Bigg(
\frac{\slashed{q}+m}{q^2-m^2+i\epsilon}\nb\\
&&
\text{VAFF}^{i_1j_1\mu_1}
\frac{\slashed{q}-\slashed{k}+m}{(q-k)^2-m^2+i\epsilon}
\text{VAFF}^{i_2j_2\mu_2}
\Bigg).
\eqn 
We now use the Cutkosky rules to derive the imaginary part of this amplitude
\bqn
&&2\text{Im}\left({\cal{M}}\right)=-\int \frac{d^4q}{(2\pi)^4} {\cal{M}}_{1_{i_1j_1\mu_1}}{\cal{M}}_{1_{i_2j_2\mu_2}}\times\nb\\
&&\left(-2\pi\right)^2\Theta\left(q^0\right)\Theta\left(q^0-k^0\right)\times\nb\\
&&\delta\left(q^2-m^2\right)
\delta\left((q-k)^2-m^2\right)\nb\\
&&\text{Tr}\left(
(\slashed{q}+m)
\text{VAFF}^{i_1j_1\mu_1}
(\slashed{q}-\slashed{k}+m)
\text{VAFF}^{i_2j_2\mu_2}
\right).\nb\\
\eqn
On the other hand, the right hand side of Eq.~\ref{Eq:OpticalTheorem} is equal to
\bqn
\lb{Eq:RighHandOfOptical}
&&\int \frac{d^3q_1}{(2\pi)^32E_{q_1}}\int \frac{d^3q_2}{(2\pi)^32E_{q_2}}
\left(2\pi\right)^4\delta^4(k-q_1-q_2)\nb\\
&&\sum_{\text{spin}}|{\cal{M}}_{\text{half}}|^2,
\eqn
where the half amplitude is equal to 
\bqn
&&{\cal{M}}_{\text{half}} = {\cal{M}}_{1_{i_2j_2\mu_2}}\bar{v}^{\sigma_2}(q_2)
\text{VAFF}_{i_1j_1\mu_1}u^{\sigma_1}(q_1)
\nb\\
&&~~\nb\\
&&=~~~~~\begin{fmffile}{HalfAmplitude}
      \setlength{\unitlength}{0.55cm}
      \parbox{30mm}{\begin{fmfgraph*}(3.5,3.5)
    \fmfleft{i}
    \fmfright{o2,o1}
    \fmf{wiggly}{i,v}
    \fmf{fermion,label=$q_2$,label.side=right}{o2,v}
    \fmf{fermion,label=$q_1$}{v,o1}
    \fmflabel{?}{i}
    \marrow{a}{left}{top}{}{v,o2}
\end{fmfgraph*}}
\end{fmffile}.
\eqn
We now use the spin method to convert the spin sum of the half amplitudes into a trace
\bqn
&&\sum_{\text{spin}}|{\cal{M}}_{\text{half}}|^2 = {\cal{M}}_{1_{i_1j_1\mu_1}}{\cal{M}}_{1_{i_2j_2\mu_2}}^{\ast}\times\nb\\
&&\text{Tr}\left(
(\slashed{q_1}+m)
\text{VAFF}^{i_1j_1\mu_1}
(\slashed{q_2}-m)
\text{VAFF}^{i_2j_2\mu_2}
\right),
\eqn
and use 
\bqn
\int \frac{d^3q}{2E_q}=\int d^4q\Theta(q^0)\delta(q^2-m^2),
\eqn
to show that Eq.~\ref{Eq:RighHandOfOptical} is equal to $2\text{Im}\left({\cal{M}}\right)$ and, therefore, unitarity is preserved.

\section{One-loop renormalization of QLGT}
 \renewcommand{\theequation}{D.\arabic{equation}} \setcounter{equation}{0}
 \lb{sec:renormalizability}
In this section, we would like to show that all of the one-loop infinities of QLGT can be absorbed in its available parameters, and the theory is renormalizable to that order. We use the FeynCalc package to carry out the calculations within the Passarino-Veltman scheme \cite{1979NuPhB.160..151P} and make the scripts available in \cite{ComputerScriptsBorzou}.
For the calculations, we follow the instructions for the one-loop calculations of QED using the same computational package in \cite{2018.book.Romao}.

\paragraph{Fermion self-energy\\}
\lb{Sec:FermionSelfEnergy}
The correction to the fermion propagator is through the following diagram
\bqn
&&\begin{fmffile}{FermionSelfEnergy}
      \setlength{\unitlength}{0.55cm}
      \parbox{30mm}{\begin{fmfgraph*}(5,3)
       \fmfleft{i}
       \fmfright{o}
       \fmftop{m}
       \fmf{fermion,label=$p$,tension=1}{i,v1}
       \fmf{fermion,label=$p$,tension=1}{v2,o}
       \fmf{fermion,tension=1}{v1,v2}
       \fmf{photon,tension=0}{v1,m,v2}
\end{fmfgraph*}}
\end{fmffile} = -i\Sigma(p)=\int \frac{d^4k}{(2\pi)^4}\nb\\
&&\frac{i\delta_{ij,mn}\eta_{\mu\nu}}{k^2}\text{VAFF}^{ij\mu}\frac{i(\slashed{p}+\slashed{k}+m)}{(p+k)^2-m^2}\text{VAFF}^{mn\nu}
.
\eqn
To reduce the computational load, we rewrite the vertex in the following form
\bqn
\text{VAFF}^{ij\mu}=\frac{-ig}{4}\epsilon^{kij\mu}\gamma_k\gamma^5,
\eqn 
and use the following identity
\bqn
\delta_{ij,mn}\eta_{\mu\nu}\epsilon^{k_1ij\mu}\epsilon^{k_2mn\nu}=-6\eta^{k_1k_2},
\eqn
where $\epsilon^{kij\mu}$ is the Levi-Civita symbol.
The loop reads
\bqn
-i\Sigma(p) = iA +iB\slashed{p},
\eqn
where the two infinite parameters are
\bqn
&&A = \frac{3g^2m}{64 \pi^2}\left(1-2B_0(p^2,0,m^2)\right),\nb\\
&&B = \frac{3g^2}{128\pi^2}\Big(1+B_0(0,0,m^2)-2B_0(m^2,0,m^2)\Big).
\eqn
Here and in the rest of the paper, $A_0(\cdots)$, $B_0(\cdots)$ and $C_0(\cdots)$ are the Passarino-Veltman functions.

To see the corrections to the mass and the spinor field, we note that the exact fermion propagator is equal to
\bqn
\text{PF}_{(t)}&=&\text{PF}+ 
\text{PF}\left(-i\Sigma\right)\text{PF}  \nb\\
&&+\text{PF}\left(-i\Sigma\right)\text{PF}\left(-i\Sigma\right)\text{PF}+ \cdots\nb\\
&=&
\text{PF}\left(1+\left(-i\Sigma\right)\text{PF}_{(t)}\right).
\eqn
Therefore, the inverse of the propagators satisfy the following equation
\bqn
\lb{Eq:InvFermionPropag}
\text{PF}_{(t)}^{-1}&=&\text{PF}^{-1}+i\Sigma\nb\\
&=&-i\left(\slashed{p}-m\right)+i\Sigma.
\eqn
To absorb the two infinities of $\Sigma$, we define the renormalized parameters as
\bqn
&&\psi^r \equiv \frac{1}{\sqrt{Z_{\psi}}}\psi,\nb\\
&&m^r \equiv \frac{1}{Z_m}m,
\eqn
where $Z_{\psi/m} \equiv 1 + \delta_{\psi/m}$. Since the fermion propagator is by definition $\langle 0|\psi\bar{\psi}|0\rangle$, the renormalization is equivalent to $\text{PF}\rightarrow Z_{\psi}^{-1}\text{PF}$, and Eq.~\ref{Eq:InvFermionPropag} reads
\bqn
\text{PF}^{-1}_{(t)}=-i\left(
\slashed{p}\left(1+\delta_{\psi}-B\right)
-m^r-\left(\delta_m+\delta_{\psi}\right)m^r-A
\right).\nb\\
\eqn
To remove the infinities, we now define 
\bqn
&&\delta_{\psi} = \text{Div}(B),\nb\\
&&\delta_{\psi}+\delta_m=-\frac{1}{m^r}\text{Div}(A),
\eqn
where Div stands for the divergent component of the expression. 

\paragraph{Vacuum polarization\\}
The corrections to the Lorentz gauge field propagator are through four loop diagrams that are calculated below. Whenever applicable, we use the Feynman–'t Hooft gauge of $\xi=1$.

The first diagram to consider is the correction by a fermionic loop
\bqn
&&\begin{fmffile}{VacuumPolarization}
   \setlength{\unitlength}{0.55cm}
      \parbox{30mm}{
    \begin{fmfgraph*}(4,3)
       \fmfleft{i}
       \fmfright{o}
       \fmf{wiggly,tension=0.5,label=$k$}{i,v1} 
       \fmf{wiggly,tension=0.5,label=$k$}{v2,o}
       \fmf{fermion,left,tension=0.2,label=$p$}{v1,v2}
       \fmf{fermion,left,tension=0.2,label=$p+k$}{v2,v1}
    \end{fmfgraph*}}     
\end{fmffile} \equiv i\Pi^{ij\mu,mn\nu}_1 = \int \frac{d^4p}{(2\pi)^4}\nb\\
~\nb\\
&&
\frac{\text{Tr}\Big(
\text{VAFF}^{ij\mu}\cdot(\slashed{p}+\slashed{k}+m)\cdot\text{VAFF}^{mn\nu}\cdot(\slashed{p}+m)
\Big)}
{\big((p+k)^2-m^2\big)\big(p^2-m^2\big)}
\nb\\
~\nb\\
&&=
\frac{i g^2}{192 \pi ^2}
\text{B}_0\left(0,m^2,m^2\right)
\left(6 m^2 \eta_{k_1k_2}+k_{k_1}k_{k_2}\right) 
\epsilon ^{k_1 i j \mu}\epsilon^{k_2 m n\nu},\nb\\
\lb{eq:vacPolFermLoop}
\eqn 
where we have given a fictitious mass $\lambda$ to the gauge field and used 
\bqn
\lim_{\lambda^2\rightarrow 0}
\frac{B_0\left(\lambda^2,m^2,m^2\right)-B_0\left(0,m^2,m^2\right)}{\lambda^2}
=\frac{1}{6m^2}.
\eqn

The second diagram is the correction by two first order gauge field vertices
\bqn
&&\begin{fmffile}{VacuumPolarization2}
   \setlength{\unitlength}{0.55cm}
      \parbox{30mm}{
    \begin{fmfgraph*}(4,3)
       \fmfleft{i}
       \fmfright{o}
       \fmf{wiggly,tension=0.5,label=$k$}{i,v1} 
       \fmf{wiggly,tension=0.5,label=$k$}{v2,o}
       \fmf{wiggly,left,tension=0.2,label=$p$}{v1,v2}
       \fmf{wiggly,left,tension=0.2,label=$p+k$}{v2,v1}
    \end{fmfgraph*}}     
\end{fmffile} \equiv i\Pi^{ij\mu,mn\nu}_2 = \int \frac{d^4p}{(2\pi)^4}\nb\\
~\nb\\
&&
\text{V3A}^{ij\mu,i_1j_1\mu_1,i_2j_2\mu_2}(k,p,-p-k)
\textsf{PA}_{i_1j_1\mu_1,m_1n_1\nu_1}(p)\nb\\
&&
\text{V3A}^{m_1n_1\nu_1,m_2n_2\nu_2,mn\nu}(-p,p+k,-k)\nb\\
&&
\textsf{PA}_{m_2n_2\nu_2,i_2j_2\mu_2}(p+k)\nb\\
&&
=
\frac{4 i g^2}{9 \pi ^2} k^{\mu}k^{\nu } \delta^{ij,mn},
\eqn
which is finite and needs no counter term. 
Note that this is a modification to the longitudinal component of the Lorentz gauge field. 
In the gauge theories based on the unitary groups, the correction is always to the transverse component, and the longitudinal ghost component remains suppressed. In QLGT, however, the gauge field is not a tensor, as was discussed above. Hence, neither the longitudinal nor the transverse components represent an observable particle. Therefore, the correction to the longitudinal component of the Lorentz gauge field does not have adversarial effects. 

The third correction is from the second order vertex of the Lorentz gauge field
\bqn
\begin{fmffile}{VacuumPolarization3}
   \setlength{\unitlength}{0.55cm}
      \parbox{30mm}{
    \begin{fmfgraph*}(4,3)
       \fmfleft{i}
       \fmfright{o}
       \fmftop{m}
       \fmf{wiggly,tension=1}{i,v1}
       \fmf{wiggly,tension=1}{v1,o}
       \fmf{wiggly,right,tension=0}{v1,m,v1}
    \end{fmfgraph*}}     
\end{fmffile}
&\equiv & i\Pi^{ij\mu,mn\nu}_3 \propto \int \frac{d^4p}{(2\pi)^4}\frac{1}{p^2}\nb\\
&=&i\pi^2 A_0(0)=0.
\eqn
The loop is zero because the V4A vertex is momentum independent and can be taken out of the integral.

Finally, the last correction is from the ghost vertices
\bqn
&&\begin{fmffile}{VacuumPolarization4}
   \setlength{\unitlength}{0.55cm}
      \parbox{30mm}{
    \begin{fmfgraph*}(4,3)
       \fmfleft{i}
       \fmfright{o}
       \fmf{wiggly,tension=0.5,label=$k$}{i,v1} 
       \fmf{wiggly,tension=0.5,label=$k$}{v2,o}
       \fmf{dots,left,tension=0.2,label=$p$}{v1,v2}
       \fmf{dots,left,tension=0.2,label=$p+k$}{v2,v1}
    \end{fmfgraph*}}     
\end{fmffile} \equiv i\Pi^{ij\mu,mn\nu}_4 = -\int \frac{d^4p}{(2\pi)^4}\nb\\
~\nb\\
&&
\text{VAc}\bar{\text{c}}^{a_2 b_2, ij \mu_, a_1 b_1}(p+k)
\textsf{Pc}(p)_{a_1b_1,a_3b_3}\nb\\
&&
\text{VAc}\bar{\text{c}}^{a_3 b_3, mn \nu_, a_4 b_4}(p)
\textsf{Pc}_{a_4b_4,a_2b_2}(p+k)
\nb\\
&&
=
\frac{-i g^2}{72 \pi ^2} k^{\mu}k^{\nu } \delta^{ij,mn},
\eqn
which is finite, and as expected, has the same form as in $i\Pi^{ij\mu,mn\nu}_2$.

Out of the four corrections to the propagator of the Lorentz gauge field, only the fermionic loop contains infinities. 
To find the counterterms, we note that the only external lines in QLGT are fermions and the only possible Feynman diagrams have the following form
\bqn
&&\begin{fmffile}{FermionFermion}
   \setlength{\unitlength}{0.55cm}
      \parbox{40mm}{
    \begin{fmfgraph*}(6,3)
       \fmfleft{i1,i2}
       \fmfright{o1,o2}
       \fmf{fermion}{i1,v1,i2}
       \fmf{wiggly}{v1,v2}
       \fmfblob{0.25w}{v2}
       \fmf{wiggly}{v2,v3}
       \fmf{fermion}{o1,v3,o2}
    \end{fmfgraph*}}     
\end{fmffile}=\nb\\
~\nb\\
&&\chi^{e_1}\chi^{e_2} \epsilon_{e_1a_1b_1\sigma_1}\epsilon_{e_2a_2b_2\sigma_2}
\Bigg(
\textsf{PA}_{a_1b_1\sigma_1,a_2b_2\sigma_2}+\nb\\
&&
\textsf{PA}_{a_1b_1\sigma_1,ij\mu}\cdot i\Pi^{ij\mu,mn\nu}\cdot \textsf{PA}_{mn\nu,a_2b_2\sigma_2}
+{\cal{O}}(g^4)
\Bigg),\lb{Eq:FermionFermion}
\eqn
where 
$\chi^{e_1}\equiv \frac{-ig}{4}\bar{\varphi}\gamma^{e_1}\gamma^5\varphi$, 
and $\varphi$ stands for any of the spinors. If we use 
$\epsilon_{e_1}^{~~ab\sigma}\epsilon_{e_2 ab\sigma}= -6\eta_{e_1e_2}$, and 
$\epsilon_{ij}^{~~ab}\epsilon_{mnab}=-4\delta_{ij,mn}$,
expression above reads
\bqn
&&-6i\chi^{e_1}\chi^{e_2}\Bigg(
\frac{\eta_{e_1e_2}}{k^2}
+\frac{6 g^2}{192 \pi ^2k^4}
\text{B}_0\left(0,m^2,m^2\right)\cdot
\nb\\
&&
~~~~\left(6 m^2 \eta_{e_1e_2}+k_{e_1}k_{e_2}\right) 
+g^2\cdot \text{finite} +
{\cal{O}}(g^4)
\Bigg).
\eqn
We define the renormalized parameter as 
\bqn
&&A^{r}_{ij\mu}\equiv \frac{1}{\sqrt{Z_A}} A_{ij\mu},\nb\\
&&\xi^r \equiv \frac{1}{Z_{\xi}}\xi,
\eqn 
where
$Z_{A/\xi} \equiv 1+\delta_{A/\xi}$.
Subsequently, the propagator of the gauge field takes the following form
\bqn
\textsf{PA}(k)_{ij\mu,mn\nu} &=& 
\frac{i\delta^{ij,mn}}{k^2}
\Bigg(
	\left(1+\delta_A\right)\left(\eta_{\mu\nu}-\frac{k_{\mu}k_{\nu}}{k^2}\right)\nb\\
	&&+\left(1+\delta_{\xi}+\delta_A\right)\xi \frac{k_{\mu}k_{\nu}}{k^2}
\Bigg).
\eqn
Since $\delta_A$ and $\delta_{\xi}$ are of order $g^2$, only the first term in the parentheses in Eq.~\ref{Eq:FermionFermion} receives a correction from them while the correction to the rest of the terms are of the order of $g^4$ or higher and can be neglected. 
A straightforward calculation shows that the infinities can be removed if we choose
\bqn
&&\delta_A+\frac{1}{3}\delta_{\xi} = -\frac{3g^2m^2}{16\pi^2k^2}\text{Div}\left(B_0(0,m^2,m^2)\right),\nb\\
&&\delta_{\xi}=\frac{3g^2}{32\pi^2}\text{Div}\left(B_0(0,m^2,m^2)\right).
\eqn
 
We would like to emphasize that in Eq.~\ref{eq:vacPolFermLoop}, we showed that the corrections to the vacuum polarization has a form different than the one in the tree level. This is unlike any of the Yang-Mills theories of the standard model. The reason we could absorb this infinite correction by the available parameters of the theory was that the gauge field did not have an external line and the only possible Feynman diagram of the theory was given in Eq.~\ref{Eq:FermionFermion} and the fortunate fact that the contraction of two Levi-Civita symbols is proportional to the six-dimensional internal metric of the homogeneous Lorentz group.

\paragraph{Ghost self-energy\\}
The ghost propagator receives a corrections from the following loop
\bqn
&&\begin{fmffile}{GhostSelfEnergy}
      \setlength{\unitlength}{0.55cm}
      \parbox{30mm}{\begin{fmfgraph*}(5,3)
       \fmfleft{i}
       \fmfright{o}
       \fmftop{m}
       \fmf{ghost,label=$p$,tension=1}{i,v1}
       \fmf{ghost,label=$p$,tension=1}{v2,o}
       \fmf{dots,tension=1}{v1,v2}
       \fmf{wiggly,tension=0}{v1,m,v2}
\end{fmfgraph*}}
\end{fmffile}=-i\Sigma_{ij,mn}=\int\frac{d^4k}{(2\pi)^4}\nb\\
&&
\textsf{PA}_{k_1l_1\sigma_1,k_2l_2\sigma_2}(k)\text{VAc}\bar{\text{c}}^{a_1 b_1, k_1l_1\sigma_1, ij}(p+k)\nb\\
&&\textsf{Pc}(p)_{a_1b_1,a_2b_2}(p+k)\text{VAc}\bar{\text{c}}^{mn, k_2l_2\sigma_2, a_2b_2}(p)
\nb\\
&&
=-\frac{ig^2m^2}{16\pi^2}\delta_{ij,mn}\left(1-B0(m^2,0,0)\right)
.
\eqn
The correction to the inverse of the ghost propagator can be found in the same way as for the fermion self-energy above and reads
\bqn
\textsf{Pc}(p)^{-1}_{(t)ij,mn} =
-ip^2\delta_{ij,mn} +i\Sigma_{ij,mn}
\eqn

Since the ghost field has no mass, we only define one renormalized parameter as
\bqn
c^{_rij} \equiv \frac{1}{\sqrt{Z_c}}c^{ij},
\eqn
where $Z_c\equiv 1 + \delta_c$. To remove the infinity we further assume that 
\bqn
\delta_c \equiv -\frac{g^2}{16\pi^2}\text{Div}\left(B0(m^2,0,0)\right).
\eqn

\paragraph{Fermion vertex renormalization\\}
The correction to the fermion vertex is from the following two diagrams

\bqn
&&~~~~\begin{fmffile}{FermionVertex}
      \setlength{\unitlength}{0.55cm}
      \parbox{35mm}{
\begin{fmfgraph*}(5,4)
  \fmfright{i1,i2}
  \fmfleft{o}
  \fmf{fermion,label=$p_1$,tension=1.5}{i1,v1}
  \fmf{fermion}{v1,v}
  \fmf{fermion}{v,v2}
  \fmf{wiggly,label=$p_1-p_2$,tension=1.5}{o,v}
  \fmf{fermion,label=$p_2$,label.side=right,tension=1.5}{v2,i2}
  \fmf{wiggly}{v2,v1}
        \fmflabel{$ab\sigma$}{o}
\end{fmfgraph*}
}
\end{fmffile}=\Gamma^{ab\sigma}_1\nb\\
&&=\int\frac{d^4k}{(2\pi)^4}\text{VAFF}^{ij\mu}\cdot\textsf{PF}(p_2-k)
\cdot\text{VAFF}^{ab\sigma}\nb\\
&&\cdot\textsf{PF}(p_1-k)\cdot\text{VAFF}^{mn\nu} \textsf{PA}(k)_{ij\mu,mn\nu}\nb\\
&&=A\epsilon^{kab\sigma}q_k\gamma^5+B\epsilon^{kab\sigma}\gamma_k\gamma^5,
\eqn
where $q \equiv p_1-p_2$, and
\bqn
A &=& 
-\frac{3 g^3 m}{128 \pi^2}
   C_0\left(m^2,m^2,q^2,m^2,0
   ,m^2\right) ,
\nb\\
B &=&
\frac{3 g^3}{512 \pi ^2} \Bigg(-3
   B_0\left(q^2,m^2,m^2\right)+4
   B_0\left(m^2,0,m^2\right)\nb\\
   &+&\left(6
   m^2-2 q^2\right)
   C_0\left(m^2,m^2,q^2,m^2,0
   ,m^2\right)-1\Bigg).
\eqn

\bqn
&&~~~~\begin{fmffile}{FermionVertex2}
      \setlength{\unitlength}{0.55cm}
      \parbox{35mm}{
\begin{fmfgraph*}(5,4)
  \fmfright{i1,i2}
  \fmfleft{o}
  \fmf{fermion,label=$p_1$,tension=1.5}{i1,v1}
  \fmf{wiggly}{v1,v}
  \fmf{wiggly}{v,v2}
  \fmf{wiggly,label=$p_1-p_2$,tension=1.5}{o,v}
  \fmf{fermion,label=$p_2$,label.side=right,tension=1.5}{v2,i2}
  \fmf{fermion}{v1,v2}
        \fmflabel{$ab\sigma$}{o}
\end{fmfgraph*}
}
\end{fmffile}=\Gamma^{ab\sigma}_2\nb\\
&&=\int \frac{d^4k}{(2\pi)^4}
\text{VAFF}^{ij\mu}\cdot\textsf{PF}(k)
\cdot\text{VAFF}^{mn\nu}\nb\\
&&~~~~
\textsf{PA}_{ij\mu,i_1j_1\mu_1}(p_2-k)\textsf{PA}_{mn\nu,m_1n_1\nu_1}(p_1-k)
\nb\\
&&~~~~
\text{V3A}^{m_1n_1\nu_1,ab\sigma,i_1j_1\mu_1}(p_1-k,p_2-p_1,k-p_2)\nb\\
&&=\text{finite}.
\eqn

The expression for the second diagram is finite but rather lengthy and can be found in the online repository in \cite{ComputerScriptsBorzou}. Also, to reduce the computation load, we have used the simplifications that were described in Sec.~\ref{Sec:FermionSelfEnergy} as well as $\gamma^5\cdot\gamma^5=1$, and $\gamma^5\cdot\gamma^k=-\gamma^k\cdot\gamma^5$. 
Since $C_0$ is finite, only $B$ in the first diagram contains infinities. 
Therefore, the divergent correction to the fermion vertex reads
\bqn
\text{Div}\left(\Gamma^{ab\sigma}\right) 
&=&\text{Div}\left(B\right) \epsilon^{lab\sigma}\gamma_l\gamma^5\nb\\
&=&4ig^{-1}\text{Div}\left(B\right)\text{VAFF}^{ab\sigma},
\eqn
and is proportional to the bare vertex. Hence, we can remove it by renormalizing the coupling constant 
\bqn
&&g^r \equiv \frac{1}{Z_g} g,
\eqn
with $Z_g \equiv 1 + \delta_g$, and  choosing $\delta_g$ such that
\bqn
\Bigg(4ig^{-1}\text{Div}\left(B\right)
+\delta_g\Bigg)\text{VAFF}^{ab\sigma}=0.
\eqn

\paragraph{The rest of infinities\\}
By now, we have used all of the possible parameters of QLGT to remove the infinities. On the other hand, the other three vertices in Sec.~\ref{Sec:Vertices} also receive infinite corrections from the relevant loops. In this section, we would like to show that the gauge symmetry of QLGT implies three restrictions on the coefficients of the terms in the Lagrangian that removes the rest of infinities by the choices that we have made so far for the renormalized parameters. 

Inserting all of the renormalized parameters, the total Lagrangian reads
\bqn
\lb{Eq:RenormalizedTotalLagrangian}
{\cal{L}}^r_{\text{total}} &=&
\frac{i}{2}Z_{\psi}e_i^{~\mu}\bar{\psi}^r\gamma^i \partial_{\mu}\psi^r
-
\frac{i}{2}Z_{\psi}e_i^{~\mu}\bar{\psi}^r\overleftarrow{\partial}_{\mu}\gamma^i \psi^r\nb\\
&-&Z_mZ_{\psi}m\bar{\psi}^r\psi^r-\frac{ig}{4}Z_gZ_{\psi}Z_{A}^{\frac{1}{2}}A_{ij\mu}\epsilon^{lij\mu}\bar{\psi}\gamma_l\gamma^5\psi\nb\\
&+&
Z_A{\cal{L}}_{A^2}
+
Z_g Z_{A}^{\frac{3}{2}} g{\cal{L}}_{A^3}
+
Z_g^2 Z_A^2 g^2{\cal{L}}_{A^4}\nb\\
&-&Z_c
\bar{c}^{ij}\partial^{\mu}\left(\partial_{\mu}c_{ij}\right)\nb\\
&+&Z_cZ_gZ_{A}^{\frac{1}{2}} g\bar{c}^{ij}\partial^{\mu}\left(A_{i~\mu}^{~k}c_{kj}+A_{j~\mu}^{~k}c_{ik}\right),
\eqn
where ${\cal{L}}_{A^n}$ is the part of the gauge field Lagrangian containing $n$ fields.
From this renormalized Lagrangian, we can derive the corrections to the three vertices that were not directly discussed. 

To validate these corrections, we write the total renormalized Lagrangian with unknown coefficients
\bqn
\lb{Eq:RenormalizedTotalLagrangian2}
{\cal{L}}^r_{\text{total}} &=&
\frac{i}{2}Z_1e_i^{~\mu}\bar{\psi}\gamma^i \partial_{\mu}\psi
-
\frac{i}{2}Z_1e_i^{~\mu}\bar{\psi}\overleftarrow{\partial}_{\mu}\gamma^i \psi\nb\\
&-&Z_2
m\bar{\psi}\psi-\frac{ig}{4}Z_3A_{ij\mu}\epsilon^{lij\mu}\bar{\psi}\gamma_l\gamma^5\psi\nb\\
&+&Z_4
{\cal{L}}_{A^2}
+
gZ_5{\cal{L}}_{A^3}
+
g^2Z_6{\cal{L}}_{A^4}\nb\\
&-&
Z_7\bar{c}^{ij}\partial^{\mu}\left(\partial_{\mu}c_{ij}\right)
\nb\\
&+&gZ_8\bar{c}^{ij}\partial^{\mu}\left(A_{i~\mu}^{~k}c_{kj}+A_{j~\mu}^{~k}c_{ik}\right).
\eqn
We note that the renormalized theory has to be invariant under the homogeneous Lorentz transformations. This means that the following three equations should be satisfied
\bqn
\frac{Z_3}{Z_1} = \frac{Z_5}{Z_4} = \frac{Z_8}{Z_7} = \sqrt{\frac{Z_6}{Z_4}}.
\eqn
By comparison with Eq.~\ref{Eq:RenormalizedTotalLagrangian}, we can see that our choices for the infinities meet the enforced conditions.

%
%
%
%

\section{Un-compacted equations of the plane wave analysis}
 \renewcommand{\theequation}{E.\arabic{equation}} \setcounter{equation}{0}
 \lb{App:PlaneWave}
The components of Eq.~\ref{Eq:PlaneWaveSpaceFix} after using $k^{\mu}=(k,0,0,k)$ read 
\bqn
\lb{Eq:SpaceTimeGaugeFixing}
&&\Sigma_{22}=-\Sigma_{11},\nb\\
&&\Sigma_{(01)}=-\Sigma_{(13)},\nb\\
&&\Sigma_{(03)}=-\frac{1}{2}\left(\Sigma_{00}+\Sigma_{33}\right),\nb\\
&&\Sigma_{(02)}=-\Sigma_{(23)}.
\eqn

Equations~\ref{Eq:tetradPosInMomentumSpace},~\ref{Eq:MasslessMomentum}, and \ref{Eq:PlaneWaveSpaceFix} after using $k^{\mu}=(k,0,0,k)$ indicate that 
\bqn
\lb{Eq:SolutionToTetPos}
&&\varepsilon_{013}=-\varepsilon_{010},~~
\varepsilon_{023} = - \varepsilon_{020},~~
\varepsilon_{033}=-\varepsilon_{030},\nb\\
&&\varepsilon_{123}=-\varepsilon_{120},~~\varepsilon_{133}=-\varepsilon_{130},~~\varepsilon_{233}=-\varepsilon_{230},\nb\\
&&\varepsilon_{121}=\varepsilon_{122}=\varepsilon_{031}=\varepsilon_{032}=0,\nb\\
&&\varepsilon_{012}=\varepsilon_{021},~~\varepsilon_{131}=\varepsilon_{011},
~~\varepsilon_{232}=\varepsilon_{022},\nb\\
&&\varepsilon_{132}=\varepsilon_{021}=\varepsilon_{231},\nb\\
&&\Sigma_{32}=\frac{i}{k}\varepsilon_{233},
~~~~~~~~~~~~~~\Sigma_{01}=\frac{i}{k}\varepsilon_{010},\nb\\
&&\Sigma_{02}=\frac{i}{k}\varepsilon_{020},
~~~~~~~~~~~~~~\Sigma_{11}=\frac{i}{k}\varepsilon_{011},\nb\\
&&\Sigma_{22}=\frac{i}{k}\varepsilon_{022},
~~~~~~~~~~~~~~\Sigma_{12}=\frac{i}{k}\left(\varepsilon_{021}+\varepsilon_{120}\right),\nb\\
&&\Sigma_{21}=\frac{i}{k}\left(\varepsilon_{021}-\varepsilon_{120}\right),
~~\Sigma_{30}+\Sigma_{33}=\frac{i}{k}\varepsilon_{033},\nb\\
&&\Sigma_{03}+\Sigma_{00}=\frac{i}{k}\varepsilon_{030},
~~~~~~\Sigma_{31}=\frac{i}{k}\varepsilon_{133}.
\eqn
It is interesting to note that the first two lines of equations above are in agreement with Eq.~\ref{Eq:TransverseWave}.
Also, we would like to mention that the linearized Eq.~\ref{Eq:SolutionToTetPos} is to the first order of perturbation invariant under both of the symmetry transformations of LGT. 
This is very important, since, for example, $\Sigma_{31}$ is not a physical mode because it is equal to $\frac{i}{k}\epsilon_{133}$, and an appropriate transformation can remove the latter. This conclusion was not possible if equality was lost after the transformation.

\bibliographystyle{spphys}       
\bibliography{Ref}   

\end{document}